%% file: main.tex
\documentclass[format=acmsmall, review=false]{acmart}

\usepackage{acm-ec-25}
\usepackage{booktabs} % For formal tables
\usepackage[ruled]{algorithm2e} % For algorithms

\SetAlFnt{\small}
\SetAlCapFnt{\small}
\SetAlCapNameFnt{\small}
\SetAlCapHSkip{0pt}
\IncMargin{-\parindent}

\setcitestyle{acmnumeric}
\title[Adaptivity and Robustness in DeFi Lending]{AgileRate: Bringing Adaptivity and Robustness to DeFi Lending Markets}

\author{Mahsa Bastankhah}
\affiliation{%
    \institution{Princeton University}
    \city{Princeton}
    \state{NJ}
    \country{USA}
}
\email{mhs.bastankhah@princeton.edu}

\author{Viraj Nadkarni}
\affiliation{%
    \institution{Princeton University}
    \city{Princeton}
    \state{NJ}
    \country{USA}
}
\email{viraj@princeton.edu}

\author{Xuechao Wang}
\affiliation{%
    \institution{Hong Kong University of Science and Technology (Guangzhou)}
    \city{Guangzhou}
    \country{China}
}
\email{xuechaowang@hkust-gz.edu.cn}

\author{Pramod Viswanath}
\affiliation{%
    \institution{Princeton University}
    \city{Princeton}
    \state{NJ}
    \country{USA}
}
\email{pramodv@princeton.edu}

\input{macro}

\begin{document}

% Abstract. Note that this must come before \maketitle.
\begin{abstract}
Decentralized Finance (DeFi) has revolutionized lending by replacing intermediaries with algorithm-driven liquidity pools. However, existing platforms like Aave and Compound rely on static interest rate curves and collateral requirements that struggle to adapt to rapid market changes, leading to inefficiencies in utilization and increased risks of liquidations. In this work, we propose a dynamic model of the lending market based on evolving demand and supply curves, alongside an adaptive interest rate controller that responds in real-time to shifting market conditions. Using a Recursive Least Squares algorithm, our controller tracks the external market and achieves stable utilization, while also controlling default and liquidation risk. We provide theoretical guarantees on the interest rate convergence and utilization stability of our algorithm. We establish bounds on the system's vulnerability to adversarial manipulation compared to static curves, while quantifying the trade-off between adaptivity and adversarial robustness.
We propose two complementary approaches to mitigating adversarial manipulation: an algorithmic method that detects extreme demand and supply fluctuations and a market-based strategy that enhances elasticity, potentially via interest rate derivative markets.
Our dynamic curve demand/supply model demonstrates a low best-fit error on Aave data, while our interest rate controller significantly outperforms static curve protocols in maintaining optimal utilization and minimizing liquidations.

\end{abstract}

% Title page for title and abstract only.
%\begin{titlepage}

\maketitle
\settopmatter{printfolios=true}

% Optionally include a table of contents
%\vspace{1cm}
%\setcounter{tocdepth}{2} % adjust to 1 if desired
%\tableofcontents

%\end{titlepage}

% Paper body
\input{introduction}

\input{Problem_Formulation}

\input{protocol}
\input{Adversarial_analysis}

\input{evaluation}

\input{conclusion}
\bibliographystyle{ACM-Reference-Format}
\bibliography{references}

% Appendix
\appendix
\input{appendix}

\end{document}

%% file: macro.tex
\usepackage[utf8]{inputenc}
\usepackage{graphicx}
\usepackage{hyperref} % For 
\usepackage{diagbox}
\usepackage{mathtools}
\usepackage{multirow}

\usepackage{graphicx}

\usepackage{algorithm}
\usepackage{algpseudocode}

\usepackage{subcaption}

\usepackage{bbm}

\usepackage{hyperref}       % hyperlinks
\usepackage{url}            % simple URL typesetting
\usepackage{booktabs}       % professional-quality tables
\usepackage{amsfonts}       % blackboard math symbols
\usepackage{nicefrac}       % compact symbols for 1/2, etc.
\usepackage{microtype}      % microtypography
\usepackage{xcolor}         % colors
\usepackage{bm}
\usepackage{xspace}
\usepackage{appendix} % Optional: helps with appendix formatting
\usepackage{wrapfig} % Add this in the preamble
\usepackage{amsmath} % for math environments and more

\usepackage{prettyref}
\newrefformat{fig}{Figure~[\ref{#1}]}
\newrefformat{app}{Appendix~[\ref{#1}]}
\newrefformat{eq}{Equation~[\ref{#1}]}

\usepackage{float}
\usepackage{caption}
\usepackage{subcaption}

\newcommand{\indicator}[1]{\mathbbm{1}_{#1}}

\newcommand{\colfacN}{\ensuremath{c}\xspace}
\newcommand{\ut}{\ensuremath{U_t}\xspace}
\newcommand{\uS}[1]{\ensuremath{U_{#1}}\xspace}
\newcommand{\rt}{\ensuremath{r_t}\xspace}
\newcommand{\rS}[1]{\ensuremath{r_{#1}}\xspace}
\newcommand{\liqthrshN}{\ensuremath{LT}\xspace}
\newcommand{\liqIncntive}{\ensuremath{LI}\xspace}
\newcommand{\lt}{\ensuremath{L_t}\xspace}
\newcommand{\lS}[1]{\ensuremath{L_{#1}}\xspace}

\newcommand{\rolt}{\ensuremath{r_{o}^l}\xspace}
\newcommand{\robt}{\ensuremath{r_{o}^b}\xspace}

\newcommand{\bt}{\ensuremath{B_t}\xspace}

\newcommand{\bS}[1]{\ensuremath{B_{#1}}\xspace}
\newcommand{\ct}{\ensuremath{C_t}\xspace}

\newcommand{\cS}[1]{\ensuremath{C_{#1}}\xspace}
\newcommand{\pt}{\ensuremath{p_t}\xspace}
\newcommand{\pS}[1]{\ensuremath{p_{#1}}\xspace}

\newcommand{\assetOne}{\ensuremath{\mathcal{A}_l}\xspace}
\newcommand{\assetTwo}{\ensuremath{\mathcal{A}_c}\xspace}

\newcommand{\userDefault}[2]{\ensuremath{\pi_{#1}^{#2}}\xspace}
\newcommand{\poolDefault}[1]{\ensuremath{\pi_{#1}}\xspace}
\newcommand{\userLiq}[2]{\ensuremath{\lambda_{#1}^{#2}}\xspace}

\newcommand{\protocol}{\ensuremath{\mathcal{P}}\xspace}

\newcommand{\E}[1]{\mathbb{E}\left[#1\right]}

\newcommand{\uoptimal}{\ensuremath{U^{*}}\xspace}

\newcommand{\umax}{\ensuremath{U_{\text{max}}}\xspace}

\newcommand{\rminb}{\ensuremath{r_{\text{min}}}\xspace}

\newcommand{\rmaxl}{\ensuremath{r_{\text{max}}}\xspace}

\newcommand{\ab}{\ensuremath{\mathtt{a}_t^b}\xspace}
\newcommand{\bb}{\ensuremath{\mathtt{b}_t^b}\xspace}
\newcommand{\al}{\ensuremath{\mathtt{a}_t^l}\xspace}
\newcommand{\bl}{\ensuremath{\mathtt{b}_t^l}\xspace}

\newcommand{\abn}{\ensuremath{\mathtt{a}^b}\xspace}
\newcommand{\bbn}{\ensuremath{\mathtt{b}^b}\xspace}
\newcommand{\aln}{\ensuremath{\mathtt{a}^l}\xspace}
\newcommand{\bln}{\ensuremath{\mathtt{b}^l}\xspace}

\newcommand{\abhat}{\ensuremath{\hat{\mathtt{a}}_t^b}\xspace}
\newcommand{\bbhat}{\ensuremath{\hat{\mathtt{b}}_t^b}\xspace}
\newcommand{\alhat}{\ensuremath{\hat{\mathtt{a}}_t^l}\xspace}
\newcommand{\blhat}{\ensuremath{\hat{\mathtt{b}}_t^l}\xspace}

\newcommand{\ff}{\ensuremath{\rho}\xspace}

\newcommand{\ropt}{\ensuremath{r^*}\xspace}
\newcommand{\ai}{\ensuremath{\mathcal{C}_D}\xspace}
\newcommand{\advb}{\ensuremath{\mathcal{A}^b}\xspace}
\newcommand{\advl}{\ensuremath{\mathcal{A}^l}\xspace}

\newcommand{\rateDeviation}{Rate Deviation\xspace}

\newcommand{\rateDeviationN}{\ensuremath{\mathcal{R}_t^{\protocol, \nu}(\uoptimal)}\xspace}

\newcommand{\rateDeviationS}{\ensuremath{\mathcal{R}_t^{\protocol,\nu}}\xspace}

\newcommand{\advImpact}{Adversarial Impact Deviation\xspace}

\newcommand{\advImpactN}{\ensuremath{AI^{\protocol}( \delta_b, \delta_l)}\xspace}

\newcommand{\advImpactNShort}{\ensuremath{AI^{\protocol}( \delta_b, \delta_l)}\xspace}

\newcommand{\advImpactNs}{\ensuremath{AI^{\protocol}( \delta_b, \delta_l, B, L)}\xspace}

\newcommand{\rslopeOne}{\ensuremath{R_{\text{slope1}}}\xspace}

\newcommand{\rslopeTwo}{\ensuremath{R_{\text{slope2}}}\xspace}

\newcommand{\learningController}{\ensuremath{\mathtt{LC}}\xspace}

\newcommand{\effrate}{\ensuremath{r_{\text{eff}}}\xspace}
\newcommand{\colborrowers}{\ensuremath{\mathcal{B}_t}\xspace}
\newcommand{\collenders}{\ensuremath{\mathcal{L}_t}\xspace}

\newcommand{\sigmatrns}{\ensuremath{\sigma_{\text{trns}}}\xspace}
\newcommand{\sigmanoise}{\ensuremath{\sigma_{\text{noise}}}\xspace}

%% file: introduction.tex
\section{Introduction}

 \noindent\textbf{Lending markets in DeFi}  Decentralized Finance (DeFi) has transformed lending by eliminating centralized intermediaries such as banks, replacing them with transparent, algorithm-driven liquidity pools. 
 Platforms such as Aave \cite{aaveWP} and Compound \cite{compoundWP} enable lenders to provide the capital that borrowers can access by pledging collateral.
 A key goal of these protocols is to maintain stable utilization rates, adjusting interest rates to balance supply and demand \cite{irThumbRule}.
 Low utilization results in lower interest rates to encourage borrowing, while increased utilization drives up rates to manage liquidity.
 Another critical parameter is the collateral factor, which ensures that the risk of borrowing is minimized while the markets remain attractive. Setting appropriate collateral levels requires evaluating recent price behavior of the collateral asset and the lender's risk tolerance \cite{riskColDeFi}.
 
 \noindent\textbf{Current approaches} Current DeFi platforms determine interest rates based on a static function of utilization \cite{aaveIR,compoundIR}. This approach uses utilization to represent supply/demand dynamics, market risk, and attractiveness, relying on a manually set, arbitrary interest rate curve. Additionally, collateral factors in these markets are established through a thorough process involving community proposals and reviews \cite{aaveGov,compoundGov}, which are voted on every few months or so.

\begin{figure}[ht]
    \centering
    \begin{subfigure}[t]{0.48\textwidth}
        \centering
        \includegraphics[height=0.7\textwidth]{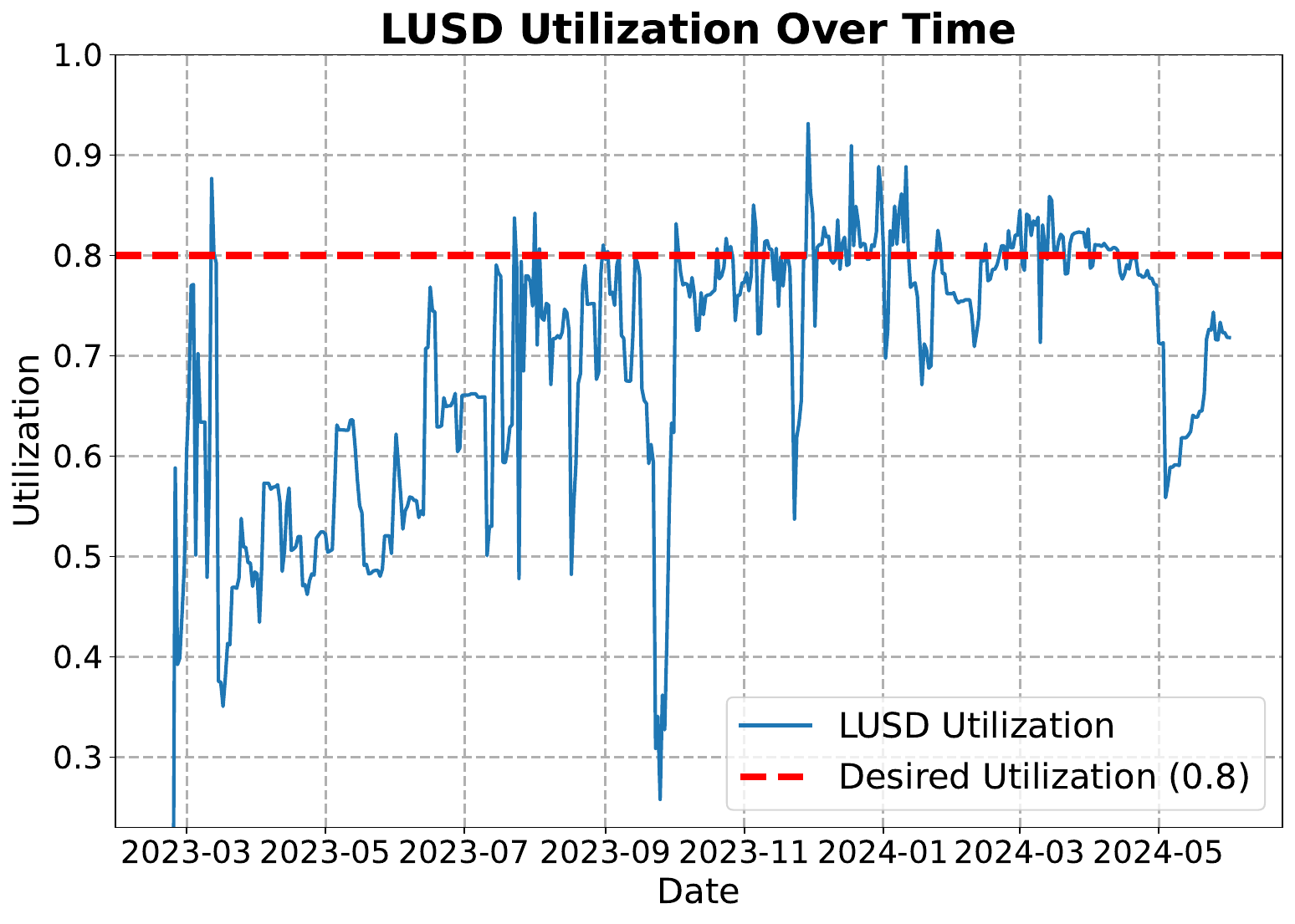}
        \caption{Lending platforms fail to adapt to market demand and supply changes}
        \label{fig:utilization}
    \end{subfigure}
    \hfill
    \begin{subfigure}[t]{0.48\textwidth}
        \centering
        \includegraphics[height=0.7\textwidth]{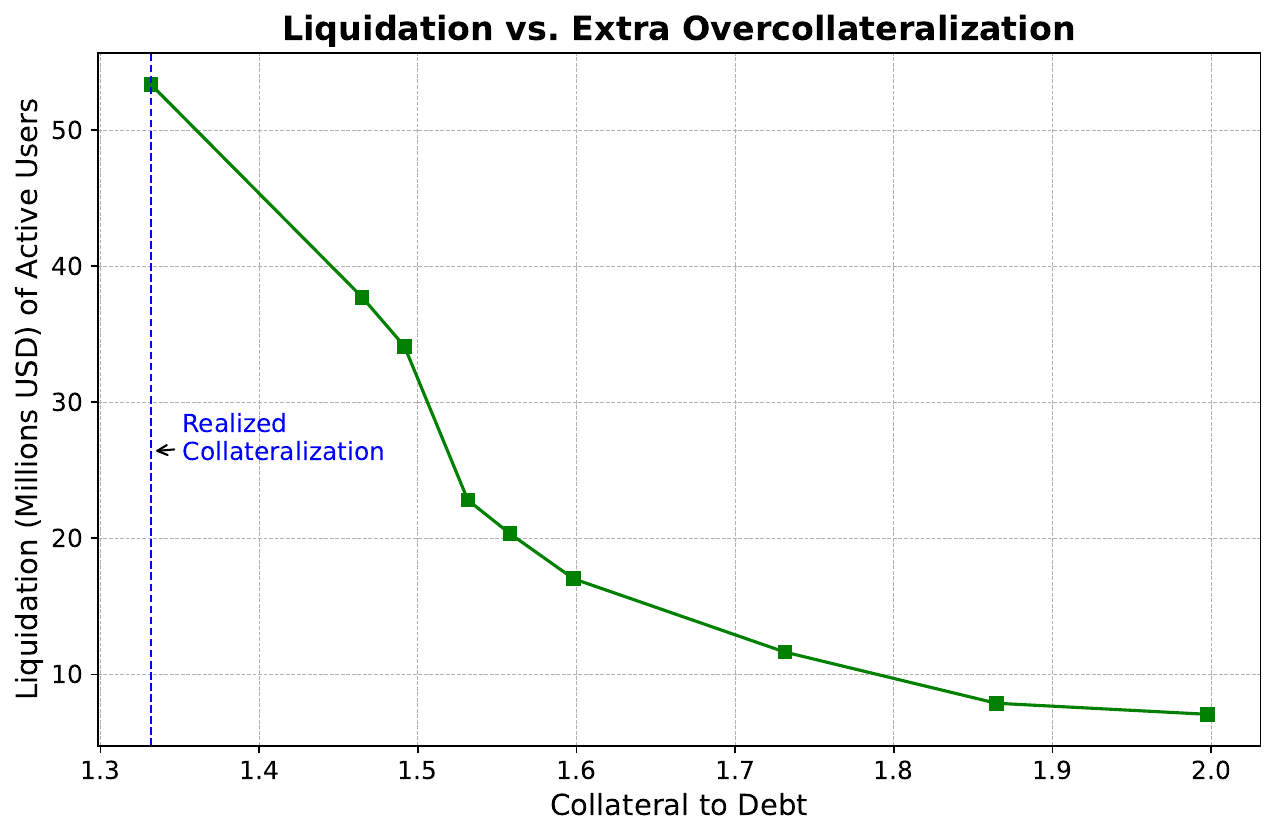}
        \caption{A slight increase in the collateral ratio could have reduced liquidations}
        \label{fig:avoidable_liquidation}
    \end{subfigure}
    \label{fig:combined_figures}
\end{figure}

\noindent\textbf{Challenges in adapting to market conditions} Current DeFi systems are slow to adapt to rapid market changes, leading to potential losses and increased risks due to delayed parameter adjustments, particularly during major price fluctuations. For instance, between August 1 and August 6, 2024, Aave ETH V3 experienced  over \$116M in liquidations due to a significant price drop in ETH. Among the affected users, only 45\% (\$53M) actively increased their collateral just before being liquidated. It turns out that a 15\% increase in the over-collateralization ratio could have nearly halved the number of liquidations for these users (see Figure \ref{fig:avoidable_liquidation}). However, Aave's slow response in adjusting collateral ratios and liquidation thresholds failed to avert these outcomes. Additionally, Aave has faced challenges in maintaining optimal utilization levels during rapid, unforeseen market changes, particularly when demand spikes due to new yield farming opportunities or similar events. For example, Figure \ref{fig:utilization} illustrates the utilization of the Aave LUSD-ETH pool during a period of significant volatility in LUSD demand, driven by unstable yield farming opportunities. Although the protocol aims to maintain a target utilization of 0.8, it frequently fails to achieve this target, with deviations persisting for extended periods.

\noindent\textbf{Modeling markets and adaptive lending pools} The first challenge in designing a better lending protocols is to model the existing population of DeFi lenders and borrowers. We start this work by proposing a model based on changing demand and supply curves with noise (\prettyref{sec:model}). We define an optimal interest rate controller to be the one that sets the rate exactly equal to the equilibrium rate that realizes the desired utilization given the current market conditions. Our goal is to design an optimally adaptive lending protocol for this model, and to quantify the agility with which the protocol can change based on external market conditions. 

\noindent\textbf{Adaptivity versus robustness} Introducing adaptivity into a protocol can increase its vulnerability to manipulation \cite{msGauntletLending23}, as an adaptive protocol adjusts based on recent market activities, learning from the trades made by borrowers and lenders. However, an adversary can exploit this by distorting the historical trade data, making the protocol respond to false shifts in demand or supply. Therefore, it is crucial to quantify the tradeoff between adaptivity and adversarial robustness before deploying any new lending protocol. Specifically, we assess how much the interest rate can be manipulated when an adversary controls a portion of the total demand and supply. We suggest two solutions each operating in a different scale, to mitigate the adversarial manipulations.

The following section summarizes our results. We set the broader research context in which we make these contributions in \prettyref{app:related}.

\subsection{Our Contributions}

\noindent\textbf{Dynamic model and evaluation metrics:}(\prettyref{sec:model}) We propose a dynamic model of the lending market via evolving demand and supply curves for the lent asset. We introduce three core metrics to evaluate DeFi lending platforms, focusing on targeting a specific utilization, liquidation/default risk, and the robustness of the protocol to adversarial manipulations.

\noindent\textbf{Agile interest rate controller:} (\prettyref{sec:interestrate-controller}) We propose an adaptive interest rate controller that adjusts dynamically based on real-time market conditions, ensuring a desired stable utilization in presence of changing environments. The market conditions are estimated in an online manner, using a variant of the Recursive Least Squares algorithm. We provide theoretical guarantees on the controller's convergence speed to the target interest rate. Additionally, we outline how the algorithm can be augmented with a risk minimization controller, which keeps defaults and liquidations below a desired threshold.

\noindent\textbf{Bounds on adversarial robustness:} (\prettyref{sec:adversarial}) We establish theoretical bounds on the manipulability of the interest rate as set by our adaptive protocol and compare it with a bound for a static curve. We show that adaptive protocols can be manipulated arbitrarily in presence of inelastic demand/supply of loans, while the same remains bounded for static protocols.

\noindent\textbf{Enhancing the adversarial robustness:} (\prettyref{sec:mitigate_adv}) We propose two complementary methods to mitigate adversarial manipulation, each addressing the issue from a different perspective. The first is an algorithmic approach based on robust recursive least squares, designed to detect and filter out improbable demand and supply patterns. The second focuses on structural market adjustments that encourage greater user elasticity, strengthening the lending pool’s resilience against adversarial manipulation.

\noindent\textbf{Evaluation:} (\prettyref{sec:evaluation}) We justify our dynamic demand/supply curve model by fitting it on data from lending pools on Aave, obtaining a $5\%$ best-fit error. We complement the theoretical guarantees of previous sections with empirical tests of the convergence and robustness of our algorithm.

%% file: Problem_Formulation.tex
\section{Problem formulation}\label{sec:model}

\subsection{Market participants and pool mechanics}\label{subsec:market-actors}
DeFi lending platforms consist of three main participants: borrowers, lenders, and liquidators, who interact via a smart contract, denoted as \protocol. These interactions occur in discrete time slots, each corresponding to a block.

\noindent\textbf{Borrowers} Borrowers secure loans by providing collateral in asset \(\assetTwo\) to borrow asset \(\assetOne\) from a pooled system. The loan amount for borrower \(i\) at time \(t\) is \(\bt(i)\), with total borrowings \(\bt\), and their collateral is \(\ct(i)\), with total collateral \(\ct\). To open a borrowing position, borrowers must satisfy $\frac{\bt(i)}{\ct(i)\pt} < \colfacN_t < 1$, where \(\pt\) is the price of \(\assetTwo\) relative to \(\assetOne\), and \(\colfacN_t\) is the protocol-defined collateral factor. The protocol applies a global interest rate \(\rt\) to all borrowings, increasing \(\bt\) over time.

\noindent\textbf{Lenders} Lenders deposit \(\assetOne\) to earn interest from borrowers. Unlike many platforms that reserve interest for risk funds and token rewards \cite{aaveWP, compoundWP}, this protocol distributes all accrued interest to lenders, making them directly bear default risks. Let \(\lt(i)\) be lender \(i\)'s deposit and \(\lt\) be the total deposits, with utilization rate \(\ut = \frac{\bt}{\lt}\). Lenders earn interest at \(\rt\) on borrowed funds, and in case of default, losses are proportional to their deposits.

\noindent\textbf{Liquidators} To prevent risks from falling collateral prices, a liquidation mechanism allows third-party liquidators to intervene when a borrower’s loan-to-value (LTV) ratio exceeds the liquidation threshold \(\liqthrshN_t < 1\). A liquidator repays part of the borrower’s debt in \(\assetOne\), claims a corresponding amount of \(\assetTwo\), and earns an incentive fee \(\liqIncntive_t\). Liquidations continue until the LTV ratio falls below the liquidation threshold \(\liqthrshN_t\).

\noindent\textbf{Protocol (\protocol)}
The protocol is encoded in the smart contract and dictates the pool parameters \(\{\rt, \colfacN_t, \liqthrshN_t, \liqIncntive_t\}\) for each timeslot, following predetermined rules. The smart contract also provides an interface for users to interact with the pool under these conditions.

\subsection{Asset price model}

We model the system in discrete time intervals, normalized to the block time. The price of the collateral asset \(\pt\) is tracked from block to block, while the lent asset, \(\assetOne\), is assumed to be a stablecoin with minimal price fluctuation. The collateral asset price \(\pt\) follows an exogenous geometric Brownian motion with stochastic volatility \(\sigma_t\), based on the Heston model \cite{heston1993stochasticvol}.

\begin{align}\label{eq:price}
&\sigma_{t}^2 = \sigma_{t-1}^2 + \kappa (\theta - \sigma_{t-1}^2) + \xi \sigma_{t-1} \eta_{t}, \quad \eta_{t} \sim \mathcal{N}(0, 1) \\
&p_t = p_{t-1} \exp\left(\left(\mu - \frac{\sigma_{t}^2}{2}\right) + \sigma_{t}  \varepsilon_t \right), \qquad \varepsilon_t \sim \mathcal{N}(0, 1)
\end{align}

\noindent Here, \(\mu\) is the drift, \(\sigma_t\) is the stochastic volatility, \(\kappa\) is the mean-reversion rate, \(\theta\) is the long-term volatility mean, \(\xi\) represents the volatility of volatility, \(\eta_t\) is a normal random variable for the volatility process, and \(\varepsilon_t\) is a normal random variable for the asset price.

\subsection{Users' incentives}\label{subsec:user-incentive}
Each borrower (lender) $i$ is specified with their time varying demand (supply) capacity denoted by $B_t(i)$ ($L_t(i)$) and their private value \(r^b_t(i)\) (\(r^l_t(i)\)), representing the maximum (minimum) interest rate they are willing to pay (receive from) \protocol to remain engaged without seeking alternative markets. 
The rates \(r^b_t(i)\) and \(r^l_t(i)\), referred to as \emph{external interest rates} in this paper, represent the rates that alternative markets with comparable risk to \protocol would offer to borrower (or lender) \(i\). We represent the vector consisting of all borrowers (lenders) external rate by $\mathbf{\hat{r}}^b_t$ ( $\mathbf{\hat{r}}^l_t$) and their capacity vector respectively by $\mathbf{\hat{B}}_t$ ($\mathbf{\hat{L}}_t$)\\

\noindent\textbf{Truthful vs strategic} Borrower (Lender) \(i\) is considered \emph{truthful} if, at any given time \(t\) and given the interest rate \(\rt\) set by \protocol, they borrow (lend) to their maximum capacity i.e., $\bt(i)$ ($\lt(i)$) if and only if \(\rt \leq r_t^b(i)\) (\(\rt\ut \geq r_t^l(i)\)). Any borrower/lender who deviates from this strategy is considered \emph{strategic}.

\noindent\textbf{True demand/supply curve}  
The true demand curve is the total demand at interest rate $r$ when all the borrowers are truthful, this function is parameterized by the external interest rate and capacity of the borrowers collectively denoted by $\colborrowers\coloneqq \{\mathbf{\hat{B}}_t, \mathbf{\hat{r}}^b_t\}$ :
\begin{equation}\label{eq:demand-curve}
f_{}(r;\colborrowers) \coloneqq \sum_{i} \bt(i) \indicator{(r^b_t(i) \geq r)}
\end{equation}
Similarly, the true supply curve represents the total supply at an effective interest rate when all the lenders are truthful; suppliers' effective rate is the interest rate they receive on the utilized part of the pool i.e., $r_{\text{eff}}\coloneqq r\cdot U$. The supply function is parameterized by the external interest rate and capacity of the lenders collectively denoted by$\mathcal{L}\coloneqq \{\mathbf{\hat{L}}_t, \mathbf{\hat{r}}^l_t\}$ :
\begin{equation}\label{eq:supply-curve}
g_{}(\effrate; \collenders ) \coloneqq \sum_{i} \lt(i) \indicator{(r^l_t(i) \leq \effrate)} 
\end{equation}
%In this paper, the true demand and supply curves are parameterized by \(\theta_t\) and \(\omega_t\), which are subject to temporal variations, reflecting changes in user behavior.

\noindent\textbf{Utility function}
Strategic lenders and borrowers decide how much to allocate to the protocol versus external markets based on their expected returns. A strategic lender \(i\) will allocate some portion of their supply, \(\hat{L}_t(i)\), to the protocol while placing the rest in an external market offering a rate \(r^l_t(i)\). The lender’s goal is to maximize their overall returns from both sources over time. Similarly, a strategic borrower \(i\) selects how much to borrow, \(\hat{B}_t(i)\), to minimize their overall borrowing costs. Calculating these optimal decisions can be quite complex, so we simplify the utility functions in certain cases to better analyze user behavior within the protocol.

At any time \( t \), a strategic lender \( i \) allocates a portion (or potentially all) of their supply \( \hat{L}_t(i) \leq L_t(i) \) to \protocol and allocate the remaining supply, \( L_t(i) - \hat{L}_t(i) \), to an alternative external market that offers a rate \( r^l_t(i) \). The lender's objective is to maximize cumulative utility over time, derived from expected interest rates from both \protocol and the alternative market. 
 Formally, the cumulative utility over time for the lender is expressed as:
\begin{equation}\label{eq:model-lender-util}
    \mathbb{E}\left[\sum_{t=0}^\infty \left(\hat{L}_{t}(i)\,r_{t}\,\ut + \left(L_{t}(i) - \hat{L}_{t}(i)\right) \,r^l_{t}(i) \right)\right]
\end{equation}
\noindent Here, the expectation is taken over the randomness of protocol and the stochastic nature of the users' behaviour which in turn impact the pool's state and \protocol's decisions.
Similarly, a borrower \( i \) aims to choose their demand value $\hat{B}_t(i) \leq \bt(i)$ to minimize the sum of the expected interest rate paid over time, which can be formalized as:
\begin{equation}\label{eq:model-borrower-util}
    \mathbb{E}\left[\sum_{t=0}^\infty \left(\hat{B}_{t}(i)\,r_{t} + \left(B_{t}(i) - \hat{B}_{t}(i)\right) \,r^b_{t}(i) \right)\right]
\end{equation}
\noindent However, calculating these utility functions and determining the sequences $\{\hat{L}_t(i)\}_t$ and $\{\hat{B}_t(i)\}_t$ that optimize them can be highly complex and may not lead to truthful strategies. In this paper, whenever needed, we simplify these utility functions for specific protocols and user behavior scenarios.

\subsection{Protocol objectives}\label{sec:demand-supply-balancing}

We define the protocol objectives in a three-fold manner. The explicit mathematical definitions we use for these objectives are given in \prettyref{app:exp_obj}.

\subsubsection{Demand-supply balancing}\label{subsec:demand-supply-balancing}

In decentralized finance (DeFi) platforms, maintaining an optimal balance between supply and demand is crucial, with the key metric being proximity to the target utilization rate, \(\uoptimal\) \cite{aave_gauntlet,compound_gauntlet}. Low utilization implies inefficient use of deposits, while high utilization restricts withdrawals. To evaluate a protocol \(\protocol\)'s ability to maintain \(\uoptimal\), we define the optimal interest rate \(r^*_{\text{utl}}\) as the rate minimizing deviation from \(\uoptimal\), given the utilization function \(U = f(r; \mathcal{B}_0)/g(\effrate; \mathcal{L}_0)\) where \(\effrate = U \cdot r\). We introduce \emph{\rateDeviation}, denoted by \(\rateDeviationN\), as the expected deviation of the protocol’s interest rate from \(r^*_{\text{utl}}\), considering randomness in protocol execution and noise in user demand and supply. A more revenue-centric version of this balancing mechanism is defined in \prettyref{app:alt_util_objective} and \prettyref{app:revenue_max}.

\subsubsection{Adversarial robustness} 

Strategic borrowers and lenders can manipulate a protocol’s interest rate by simulating demand and supply responses to maximize personal utility. We analyze the extent of such manipulation by considering a protocol \(\protocol\) with truthful users following demand and supply curves \(f(r;\colborrowers)\) and \(g(\effrate;\collenders)\), alongside a strategic lender \advl\ and a strategic borrower \advb, who control a fraction \(\delta_l\) and \(\delta_b\) of maximum supply and demand, respectively, and have external rates \(r^l\) and \(r^b\). Defining \(r_{\text{truthful}}\) and \(r_{\text{strategic}}\) as the steady-state interest rates with and without strategic behavior, we introduce \emph{\advImpact}, denoted by \(\advImpactN\), as the supremum of the expected deviation between these rates over all external rate choices of strategic users.

\subsubsection{Liquidation and default risk control}

In peer-to-pool-to-peer lending, the primary risks are borrower defaults and liquidations, requiring dynamic adjustments of parameters such as the collateral factor \(\colfacN_t\), liquidation threshold \(\liqthrshN_t\), and liquidation incentive \(\liqIncntive_t\) based on price volatility. Defaults occur when a borrower's debt exceeds collateral value due to price drops, while liquidations happen when a price decline forces the protocol to sell collateral to restore safe loan-to-value ratios. These risks are formally defined in \prettyref{app:risk-metrics}. The protocol's ability to manage these risks can be quantified using metrics such as the expected value or 95th percentile of pool defaults and liquidations, computed with respect to the price distribution.

\subsection{Baseline}\label{subsec:baseline}
For the baseline, we consider protocols akin to Compound, which utilize a piecewise linear interest rate curve to ensure stability. These protocols dynamically adjust interest rates at each block according to the model:
\begin{equation}\label{eq:baseline-controller}
r_t = 
\begin{cases} 
 R_{\text{slope1}}\,\frac{U_t}{\uoptimal}, & \text{if } U_t \leq \uoptimal \\
 R_{\text{slope1}} + R_{\text{slope2}}\left(\frac{U_t - \uoptimal}{1 - \uoptimal}\right), & \text{if } U_t > \uoptimal
\end{cases}
\end{equation}
\noindent In contrast to our proposed approach, these platforms generally set collateral factors and other market parameters through offline simulations that attempt to forecast near-future market conditions. Parameters are selected based on simulation outcomes and are subject to decentralized governance voting. Since this phase happens in an offline and opaque manner by centralized companies, we cannot compare this aspect of their protocol with ours.

%% file: protocol.tex
\section{Protocol design}
\subsection{Interest rate controller}
\label{sec:interestrate-controller}
\noindent\textbf{Demand and supply curve} We model the overall supply (\(\lt\)) and demand (\(\bt\)) as a linear functions of the interest rate (\(\rt\)) of the previous timeslot: 

\begin{align}\label{eq:demand-model}
    B &= 
    \begin{cases} 
    \lt, & \qquad\quad \text{if } \rt < \rminb, \\
    -\ab \rt + \bb, & \qquad\quad\text{if } \rminb \le \rt \le \rmaxl, \\
    0, & \qquad\quad \text{if } \rt > \rmaxl,
    \end{cases}\\\notag
    \bS{t+1} &= \min\{B+\varepsilon_t , \lt\}\quad\quad\quad \varepsilon_t \sim \mathcal{N}(0,\nu)
\end{align}
\begin{align}\label{eq:supply-model}
    L &= 
    \begin{cases} 
    \bt, & \qquad\quad\text{if } \rt \ut < \rminb, \\
    \al \rt \ut - \bl, & \qquad\quad\text{if } \rminb \le \rt \ut \le \rmaxl, \\
    \infty, & \qquad\quad\text{if } \rt \ut > \rmaxl,
    \end{cases}\\\notag
    \lS{t+1} &= \max\{L+\varepsilon_t , \bt\}\quad\quad\quad \varepsilon_t \sim \mathcal{N}(0,\nu)
\end{align}
\noindent Where \(\ab, \bb, \al, \bl \in \mathbb{R}^+\), and \( \frac{\bl}{\al} < \rminb < \rmaxl < \frac{\bb}{\ab}\). This model assumes that within a moderate range of interest rates, demand decreases linearly with the interest rate, and supply increases linearly with the lender's \emph{effective interest rate} (\(\rt \ut\)). Outside of this moderate range, lenders and borrowers react drastically. Borrowers may attempt to take out all available funds if interest rates are too low or repay their entire debt if rates are too high. Similarly, lenders may deposit all their available funds when rates are extremely favorable or withdraw all their funds when rates are unattractive.

\noindent\textbf{Estimating the parameters} Our proposed interest rate controller employs a Recursive Least Squares (RLS) estimator with a forgetting factor to adaptively estimate the parameters \(\ab\), \(\bb\), \(\al\), and \(\bl\). The RLS with a forgetting factor is a recursive version of the least squares estimator designed to handle systems modeled by \( y_t = \mathbf{x}_t^T\boldsymbol{\theta}_t + \varepsilon_t \), where $\boldsymbol{\theta}_t$ is the state of the system that can change arbitrarily over time; And \(\varepsilon_t\) is an independent, zero-mean noise term. Given a new sample \((\mathbf{x}_t, y_t)\) at any time \(t\), the RLS estimator updates its estimate of \(\boldsymbol{\theta}_t\), denoted as \(\hat{\boldsymbol{\theta}}_t\), by minimizing the following cost function \cite{islam2019recursive, canetti1989convergence}:
\begin{equation}\label{eq:loss-func-basic}
    J_t(\boldsymbol{\theta}) = \sum_{\tau=0}^t \ff^{t-\tau}  (y_\tau - \mathbf{x}_\tau^T \boldsymbol{\theta})^2,
\end{equation}
\noindent Where \(0 < \ff < 1\) is the forgetting factor, it assigns greater weight to more recent data compared to older data. The RLS algorithm, when incorporating a forgetting factor, dynamically updates its estimates based on new data. Similar to the Kalman filter, at each iteration, the algorithm calculates an estimation of the error covariance matrix, \(\mathbf{P}_{t-1}\), which is then used to determine the gain matrix, \(\mathbf{K}_{t}\). This gain matrix adjusts the influence of the new data point on the current estimate. If the estimated error is large or the forgetting factor is low (indicating the algorithm forgets faster), recent data points have a greater impact on the updated estimation. This process ensures that the algorithm effectively blends new information with past estimates, adapting to changes in the underlying data over time.
%
% This approach is particularly useful in scenarios where we have a linear observation model but the state vector \(\boldsymbol{\theta}\) changes over time. 
%
Algorithm \ref{alg:rls-forgetting-factor} outlines the use of RLS to estimate the parameters of the demand and supply curves.

\noindent\textbf{Optimizing interest rate for desired utilization} Given the demand and supply models presented in Equations \ref{eq:demand-model} and \ref{eq:supply-model}, when the noise is small, the utilization rate can be approximated with: 
\begin{equation}\label{eq:utilization}
    \uS{t+1} \approx 
    \begin{cases} 
    1 & \text{if } \rt < \rminb, \\
    \frac{-\ab \rt + \bb}{\al \rt \ut - \bl} & \text{if } \rminb \le \rt, \text{ and } \rt\ut \ge \rmaxl, \\
    0 & \text{if } \rt \ut > \rmaxl, \\
    \end{cases}
\end{equation}
\noindent The objective of our protocol is to set the interest rate \(\rt\) such that the utilization remains close to a desired utilization \(\uoptimal\). The optimal interest rate \(\ropt_{\text{utl}}\) that achieves this desired utilization can be derived as follows: \footnote{We develop similar results for the objective of maximizing revenue instead of optimizing rate deviation. See \prettyref{app:revenue_max}}
\begin{equation}\label{eq:optimal-r-for-desired-u}
    \frac{-\ab \,\ropt_{\text{utl}} + \bb}{\al \ropt_{\text{utl}}\, \uoptimal - \bl} = \uoptimal \implies \ropt_{\text{utl}} = \frac{\bb + \bl \uoptimal}{\ab + \al (\uoptimal)^2}
\end{equation}
%The desired utilization \(\uoptimal\) can only be realized if the parameters of the models satisfy \(
%\rminb \le \ropt \le \frac{\bb}{\ab}, \text{ and } \frac{\bl}{\al} \le \ropt \uoptimal \le \rmaxl.
%\)
We have developed an interest rate controller module, detailed in Algorithm \ref{alg:optimizer}, that determines the optimal interest rate using parameters estimated from the RLS algorithm. To promote exploration when estimation error is high, the controller module samples \(\rt\) from a Gaussian distribution. The mean of this distribution is the $\ropt_{\text{utl}}$ calculated using the latest estimated parameters \(\abhat\), \(\bbhat\), \(\alhat\), \(\blhat\), and the variance is derived from the error covariance matrix calculated by the RLS algorithm.
This approach allows for more diverse data points when error is high; As the algorithm progresses and the parameter estimates become more accurate, the variance decreases, leading to less randomization. Over time, the rate converges more closely to the optimal value, ensuring more precise results.

\noindent\textbf{Theoretical guarantees on the \rateDeviation}
We consider a canonical scenario where the parameters of the demand and supply curves, initially set to certain values, change to new values \(\abn, \bbn, \aln, \bln\) at time \(t_0\). We then compare the \rateDeviation of the RLS-based controller with the baseline controller. For simplicity, we slightly abuse the notation and denote the rate deviation by \(\rateDeviationS\).

\begin{theorem}\label{theorem:rate-deviation-RLS}
    Consider using Algorithm \ref{alg:optimizer} to regulate the utilization of a resource pool, where the demand and supply functions are defined by Equations \ref{eq:demand-curve} and \ref{eq:supply-curve}, respectively. If the utilization rate $\ut > 0;\,\forall t$, then the rate deviation satisfies the following bound:
    \[
    \rateDeviationS = \mathcal{O}\left(\rho^t + \psi_t(\ff)\right),
    \]
    with probability at least $1 - \delta$, where $\limsup_{t\to\infty}\psi_t(\ff)=\frac{\nu^2 (1-\ff^N)^2}{\rho^{2N} \ln\left(\frac{1}{\rho}\right)}$ and $N$ is given by:
    \(
    N = \Theta\left(\frac{\ln 1/\delta}{\ln \min_{t}\left(\text{diam}(\mathbf{P}^b_t + \mathbf{P}^l_t)\right)}\right).
    \)
\end{theorem}
\noindent Theorem \ref{theorem:rate-deviation-RLS} demonstrates that the deviation from the optimal interest rate consists of two components: a time-dependent term, which diminishes exponentially over time after changes in the demand and supply function parameters, and a persistent bias term, which remains as long as $\ff < 1$. This theorem highlights the tradeoff between adaptivity and precision. While a smaller $\ff$ leads to quicker convergence to a stable rate, it also results in greater bias in the steady state. By selecting $\ff$ sufficiently close to 1 and allowing $t$ to be sufficiently large, \rateDeviation can be made arbitrarily small. When the parameters change infrequently, a larger \(\rho\) is preferable for greater precision. However, when the frequency of parameter changes is high, it becomes more important to prioritize adaptivity over precision to reduce overall rate deviation. 

Unlike the RLS-based algorithm, the static interest rate curve cannot adapt to changes in demand and supply functions. This lack of adaptability means that whenever \(\abn\), \(\bbn\), \(\aln\), or \(\bln\) change, the interest rate set by the static curve has a persistent non decaying and non controllable bias compared to the optimal rate.
\begin{theorem}\label{theorem:rate-deviation-aave}
Consider the baseline interest rate controller given by Equation \ref{eq:baseline-controller}. If demand follows Equation \ref{eq:demand-curve} with $\abn > 0$ and supply is fixed at \( L \), then the rate deviation \(\rateDeviationS\) is $0$ if and only if
$\Delta \coloneqq \left|R_{\text{slope1}} - \frac{\bbn - L \, \uoptimal}{\abn }\right|=0$ otherwise:
\[
\rateDeviationS \geq \frac{\Delta}{1+\frac{\abn}{L}\cdot\max\{\frac{\rslopeOne}{\uoptimal}, \frac{\rslopeTwo}{1-\uoptimal}\}}
\]
\end{theorem}
\noindent Theorem \ref{theorem:rate-deviation-aave} highlights two key points: 
1) The \rateDeviation of a static interest rate curve does not decrease over time after changes in the demand and supply parameters. 
2) There is no automatic mechanism to control the \rateDeviation hence \(\rslopeOne\) must be manually adjusted each time the parameters change. Without this adjustment, \protocol will experience a persistent error.

\noindent\textbf{Adapting the forgetting factor}\label{par:adaptive-ff}
The forgetting factor plays a crucial role in determining how quickly the RLS estimator reacts to changes in the underlying parameters. An optimally tuned \(\ff(t)\) ensures higher stability and precision when parameters change slowly, while providing traceability when they change rapidly. Paleologu et al. \cite{paleologu2008robust} propose a method to adapt the forgetting factor based on the variance of the a posterior error, defined as \(\varepsilon_t \coloneqq (y_t - \mathbf{x}_t^T \hat{\boldsymbol{\theta}}_{t})^2\). They argue that for a perfectly tuned filter, the error variance is determined solely by the observation noise i.e.,
\(\label{eq:a-post-error-relation-1}
    \text{Var}(\varepsilon_t) = \hat{\sigma}(t),
\)
where \(\hat{\sigma}(t)\) is the estimated noise variance. Using this relationship and the RLS update rules, the following expression for the adaptive forgetting factor is derived:
\(
    \ff(t) = \min \left\{ \frac{\hat{\sigma}_q(t) \hat{\sigma}(t)}{\xi + \left| \hat{\sigma}_e(t) - \hat{\sigma}(t) \right|}, \lambda_{\max} \right\},
\)
where \(q_t \coloneqq \mathbf{x}_t^T \mathbf{P}_{t-1} \mathbf{x}_t\) and \(e_t \coloneqq (y_t - \mathbf{x}_t^T \hat{\boldsymbol{\theta}}_{t-1})^2\) (the a priori error), and \(\hat{\sigma}_q(t)\), \(\hat{\sigma}_e(t)\) are their respective empirical variances.
To estimate \(\hat{\sigma}^2(t)\), the noise variance, a longer period of the a priori error signal \(e(t)\) is used through an exponentially weighted moving average:
\(
\hat{\sigma}^2(t) = \beta \hat{\sigma}^2(t-1) + (1 - \beta) e^2(t).
\)
%Averaging over a longer window reduces the contribution of filter mismatch in \(e(t)\), leaving only the noise-related component. 

\subsection{Risk parameters}
\noindent\textbf{Bounding default and liquidation}  
Beyond stabilizing interest rates to maintain utilization or maximize revenue, the protocol also aims to minimize defaults and liquidations. We analyze the conditions needed to ensure near-zero expected defaults and liquidations.

\begin{lemma}\label{lem:risk}
The expected default and liquidation at time \( t + 1 \), given the liquidation threshold \(\liqthrshN_t\) and the loan-to-value ratio \(\colfacN_t\) at time \( t \), are bounded by the following conditions:
\begin{align*}
    \E{\userDefault{t}{i}(\pS{t+1})} &\leq \Phi\left(\frac{\log(\liqthrshN_t) - \mu}{\sigma}\right) - {\exp\left(\frac{\sigma^2}{2} + \mu\right)}/{\liqthrshN_t} \cdot \Phi\left(\frac{-\mu + \log(\liqthrshN_t) - \sigma^2}{\sigma}\right)
\end{align*}
\begin{align*}
    E[\userLiq{t}{i}(\pS{t+1})] &= \frac{1}{1 - \liqthrshN_t} \Bigg( \Phi \left( \frac{\ln \left( \frac{\colfacN_t}{\liqthrshN_t} \right) - \mu + \sigma^2}{\sigma} \right) - \frac{\liqthrshN_t}{\colfacN_t} e^{\mu + \sigma^2} \Phi \left( \frac{\ln \left( \frac{\colfacN_t}{\liqthrshN_t} \right) - \mu - \sigma^2}{\sigma} \right) \Bigg)
\end{align*}
\noindent where \( \Phi \) is the cumulative distribution function of the standard normal distribution and $\mu, \sigma$ are the price distribution parameters as outlined in \ref{eq:price}.
\end{lemma}

\noindent Intuitively, the lemma presented above provides a method for adjusting the liquidation threshold \(\liqthrshN_t\) and the collateral factor \(\colfacN_t\) to minimize these risks. By setting \(\liqthrshN_t\) appropriately, the protocol can keep expected defaults below a threshold, while also ensuring that liquidations remain below a certain threshold. The relationship between the loan-to-value ratio and the liquidation threshold is key for dynamically adjusting risk parameters. This approach is based on \cite{bastankhah2024thinking}, where \(\liqthrshN_t\) serves as an upper bound for the loan-to-value ratio, helping to limit the likelihood of default.

%% file: Adversarial_analysis.tex
\section{Adversarial analysis}\label{sec:adversarial}
In this section, we analyze the impact of strategic borrowers and lenders who aim to manipulate the interest rate for their own profit. We focus on two types of controllers: an abstract learning-based interest rate controller, which includes the RLS algorithm as an instance, and a static curve-based approach. 
\subsection{Learning-based interest rate controller} 
Learning-based interest rate controllers estimate demand and supply while incorporating prior beliefs to set optimal rates. As adaptivity increases, these controllers become more vulnerable to manipulation, quickly updating beliefs based on a few manipulated samples. To quantify this impact, we analyze an adaptive learning-based algorithm, \learningController, which alternates between \emph{exploration} and \emph{exploitation} phases. During exploration, the controller detects discrepancies between observed and expected demand and supply, testing various rates to update its beliefs. Once stabilized, the controller enters exploitation, setting rates based on the learned functions. This cycle repeats as new changes are detected. We model \learningController as an auction-like mechanism. In exploration, borrowers and lenders report their minimum (or maximum) acceptable rates and quantities. \learningController infers the demand and supply curves from this data. In exploitation, it sets a single rate \(\rt\), allowing lenders with bids \(\rt\) or lower to deposit, and borrowers with bids \(\rt\) or higher to receive loans. This mimics a uniform price auction, where participants' reported bids influence the interest rate, maximizing utility based on private valuations. Participant utility depends on whether the rate falls within their acceptable range and the portion of their supply or demand allocated to the protocol or external markets. More details on this auction implementation are provided in \prettyref{app:lc_auction}.

This auction is considered truthful if strategic borrowers and lenders prefer to report their true external rates, yielding the true demand and supply curves.

\begin{proposition}\label{theorem:dynamic-incentive}
If all the lenders or borrower are infinitesimal i.e., \(\frac{\lt(i)}{\lt} \approx 0\; \forall i\in \text{Lenders}\) and \(\frac{\bt(i)}{\bt} \approx 0\; \forall i \in \text{Borrowers}\) then \learningController mechanism is truthful. However, in the general case where borrowers and lenders are non infinitesimal, the mechanism is not necessarily truthful. 
\end{proposition}
\noindent When each user (borrower or lender) is infinitesimal, their individual supply or demand has no impact on the resulting interest rate \(\rt\). Consequently, their bidding strategy does not influence the price they pay or receive and they are incentivized to report their true valuation.
However, in general cases, the situation changes. For example, when lenders are inelastic and the supply is fixed, with only borrowers participating, the auction reduces to a multi-unit demand auction with a uniform price, which is known to be non-truthful and susceptible to demand reduction \cite{econ_lecture_note, engelbrecht1998multi}. The VCG (Vickrey-Clarke-Groves) auction is known to be truthful in this context; however, it assigns different prices to different participants, which is not feasible in a peer-to-pool-to-peer setting. However, a peer-to-peer protocol \textit{can} be attached on top of a pool, subject to constraint on the peer-to-peer contracting (e.g. Morpho protocol \cite{morpho}). But whether a VCG-like truthful mechanism can be implemented using Morpho is an open question and beyond the scope of our work. 
\begin{theorem}\label{theorem:rate-manipulation}
    Consider an \learningController interest rate controller and the following setting:
    \begin{itemize}
        \item Non-strategic borrowers with a demand function of the form \(-\abn \, r + \bbn(1-\delta_b)\), and a major strategic borrower denoted by \(\advb\) with demand \(\bbn \delta_b\) and private valuation \(r^b\), satisfying \(\delta_b < 1-\frac{\abn\,r^b}{\bbn}\).
        \item Non-elastic, non-strategic lenders controlling a supply of \(L(1-\delta_l)\), and a major strategic lender denoted by \(\advl\) controlling a supply of \(L \delta_l\) with private valuation \(r^l\).
    \end{itemize}
    
    Then:
    \[
    \advImpactNShort \leq \max\left\{\frac{\bbn \delta_b}{2\abn}, \frac{\uoptimal\,\delta_l\,L}{\abn(2-\delta_l)}\right\}.
    \]
\end{theorem}
\noindent The key idea behind the proof is as follows: a dominant strategic borrower can deliberately submit a very low interest rate bid, \(\hat{r}^b\), forcing the protocol to adopt this rate in order to achieve the desired utilization level. However, as other borrowers are more elastic (i.e., \(\abn\) is larger), they become more attracted to the system at these lower rates, offsetting the strategic borrower’s influence. Therefore, when other borrowers are highly elastic, the strategic borrower cannot push the rate too low. For a strategic lender, the analysis is analogous in a symmetric way. It is worth noting that even in the presence of strategic players, \learningController still manages to set the closest possible utilization to $\uoptimal$ however with a different interest rate compared to as if the users were truthful. The key takeaway is that the elasticity of truthful users plays a crucial role in the adversary's impact on \learningController. If all truthful players are inelastic, the Adversarial Impact can become unbounded.
\subsection{Static interest rate curves}
Unlike adaptive algorithms, which are vulnerable during the learning phase, static curves are memoryless, reducing the scope for adversarial manipulation. However, strategic users can still exploit their knowledge of the fixed curve's structure. We identify a tactic called \emph{strategic withholding}, where lenders supply less than their full capacity to raise the utilization rate, thus increasing the pool's interest rate and boosting their returns. Similarly, borrowers might borrow less than their maximum capacity to lower their overall interest costs by fulfilling the remaining demand externally. This strategy was first introduced by \cite{yaish2023suboptimality}.

We derive a closed-form expression for the rate manipulation caused by strategic withholding, assuming other participants are truthful and inelastic to interest rate changes. Inelastic participants simplify the problem for adversaries, as elastic participants would adjust their behavior by depositing more or borrowing less, which could limit manipulation.

Let a strategic borrower \(\advb\), with external rate \(r^b\), control \(\delta_b\) of the total demand \(B\), and a strategic lender \(\advl\), with external rate \(r^l\), control \(\delta_l\) of the total supply \(L\). Both aim to maximize their utility by adjusting how much they participate in the pool. The full utility functions for strategic borrowers and lenders are provided in \prettyref{app:utility-static}.

We now present the Adversarial Impact caused by strategic users through the \emph{strategic withholding} strategy.
\begin{theorem}\label{theorem:static-curve-manipulation}
Given the static curve described in \ref{eq:baseline-controller}, with fixed truthful demand \((1-\delta_b)B\) and fixed truthful supply \((1-\delta_l)L\), and the remaining demand \(\delta_b B\) and supply \(\delta_l L\) controlled by strategic borrowers and lenders, the Adversarial Impact is bounded by:
\begin{equation}
    \advImpactNShort \leq \max\left\{\frac{B\delta_b \rslopeTwo}{L (1-\uoptimal)}, \frac{B\delta_l \rslopeTwo}{L(1-\delta_l)(1-\uoptimal)}\right\}
\end{equation}
\end{theorem}
The intuition behind the proof is to maximize the adversary's utility and determine the worst-case rate manipulation possible compared to their truthful behavior. Static curves limit adversarial manipulation, as their impact is bounded even with inelastic truthful users. Notably, while \rslopeOne does not affect adversarial robustness, \rslopeTwo significantly influences the likelihood of strategic withholding attacks.

\subsection{Mitigating the effects of manipulation}\label{sec:mitigate_adv}

\subsubsection{Ecosystem modification}
%\color{blue}

In Theorem \prettyref{theorem:rate-manipulation}, we showed that an adaptive interest rate in lending pools is vulnerable to adversarial manipulation via strategic withholding. The core issue arises from the delayed response of liquidity demand (or supply) to deviations from the norm, where elasticity plays a crucial role in mitigating such manipulation. Higher elasticity ensures that observed demand better reflects true market demand, reducing manipulation risks. To enhance elasticity, lending pool designers can increase exposure to alternative trading venues, deploy pools across multiple blockchain platforms with robust cross-chain bridges, and foster arbitrage opportunities. Additionally, deep derivatives markets for interest rates, such as those enabled by yield tokenization (e.g., Pendle \cite{pendle2022whitepaper}), fixed-rate lending (e.g., IPOR \cite{camas2021ipor}), and futures markets (e.g., Timeswap \cite{ngo2022timeswap}), can help sophisticated traders assess fair interest rates and respond accordingly. By allowing speculators and arbitrageurs to act on discrepancies between expected and actual rates, these mechanisms increase demand and supply elasticity, making lending protocols more resilient. A transparent valuation model for such derivatives remains an open problem and is left for future research.

\color{black}

\subsubsection{Algorithmic mitigation}
To address adversarial manipulation of the demand and supply functions, we propose a robust version of the RLS algorithm where the adaptive controller evaluates the plausibility of incoming data by comparing it to prior estimations of system dynamics. Data points that deviate significantly from expected patterns are given a lower weights in estimating the demand and supply functions.
In particular, we employ the outlier-robust recursive least squares (RLS) estimator with a forgetting factor, as introduced in \cite{zou2000recursive,kovavcevic2016robust}. This robust estimator minimizes the following modified loss function, an alternative to the standard RLS loss function with a forgetting factor in Equation \ref{eq:loss-func-basic}:
\begin{equation}\label{eq:loss-func-robust}
    J_t(\boldsymbol{\theta}) = \sum_{\tau=0}^t \ff^{t-\tau} \phi\left((y_\tau - \mathbf{x}_\tau^T \boldsymbol{\theta})^2\right),
\end{equation}
where \(\phi(.)\) is the M-estimate function. The M-estimate function is a redescending function, meaning its derivative decreases to zero as the residual \(e\) becomes very large. Refer to Appendix \ref{app:m-estimate-function} for the description of the M-estimate function we adopt.
Zou et al. show that minimizing the robust loss function in Equation \ref{eq:loss-func-robust} results in a recursive algorithm similar to the standard RLS with forgetting factor, except that the gain is updated as:
\(
\mathbf{K}_t = \frac{q(e_t) \mathbf{P}_{t-1} \mathbf{x}_t}{\ff + q(e_t) \mathbf{x}_t^T \mathbf{P}_{t-1} \mathbf{x}_t},
\)
where \(e_t \coloneqq (y_t - \mathbf{x}_t^T \hat{\boldsymbol{\theta}}_{t-1})^2\) and \(q(e) \coloneqq \frac{d\phi(e)}{de}/e\).

% \color{red}
% %give examples of such markets

% %what might be better/worse ways of derivative markets - cite tradfi sources - when does elasticity increase  and when does it decrease

% Questions -
%     %1. do we have existing derivatives market on interest rates that can boost elasticity? eg pendle,IPOR if so how do they incentivize the boost?
%     2. how do services like morpho affect the elasticities?
%     3. is there a way to maybe the split the lending pools by tenor or by interest rate level to boost elasticity?
% \color{black}

%% file: evaluation.tex
\section{Evaluation}\label{sec:evaluation}
\begin{table}[t!]
\centering
\begin{tabular}{|c|c|c|c|}
\hline
\multicolumn{4}{|c|}{Demand Curve} \\
\hline
Token & DAI & USDC & USDT \\
\hline
Average Error & $2.1\%$ & $1.3\%$ & $1.2\%$ \\
\hline
\end{tabular}
\quad
\begin{tabular}{|c|c|c|c|}
\hline
\multicolumn{4}{|c|}{Supply Curve} \\
\hline
Token & DAI & USDC & USDT \\
\hline
Average Error & $1.6\%$ & $1\%$ & $1.4\%$ \\
\hline
\end{tabular}
\vspace{10pt}
\caption{Normalized error of Compound demand and supply estimates from Feb 2024 - Feb 2025.}
\label{tab:normalized-err}
\end{table}
\subsection{Empirical validation of demand and supply models:}\label{subsec:emperical-validation} We applied our RLS-based algorithm to demand and supply data from Compound market pools, analyzing the relationship between interest rates and the demand and supply curves in 3-hour intervals. Specifically, we fitted the data (Feb 2024 - Feb 2025, RLS with $\rho=0.95$) to our demand and supply models (Equations \ref{eq:supply-model} and \ref{eq:demand-model}) and estimated the parameters \(\ab\), \(\bb\), \(\al\), and \(\bl\) over time. To evaluate the accuracy of these parameter estimates, we predicted demand and supply for the next timeslot and summarized the prediction errors in Table \ref{tab:normalized-err}. The relatively low error across the three main pools demonstrates the effectiveness of our model and methodology. More details about our experiment are provided in the Appendix \ref{app:eval-estimated-params}.
\begin{figure}[hbt!]
    \centering
    \captionsetup{justification=centering}

    \begin{subfigure}[t]{0.47\textwidth}
        \centering
        \includegraphics[width=\linewidth]{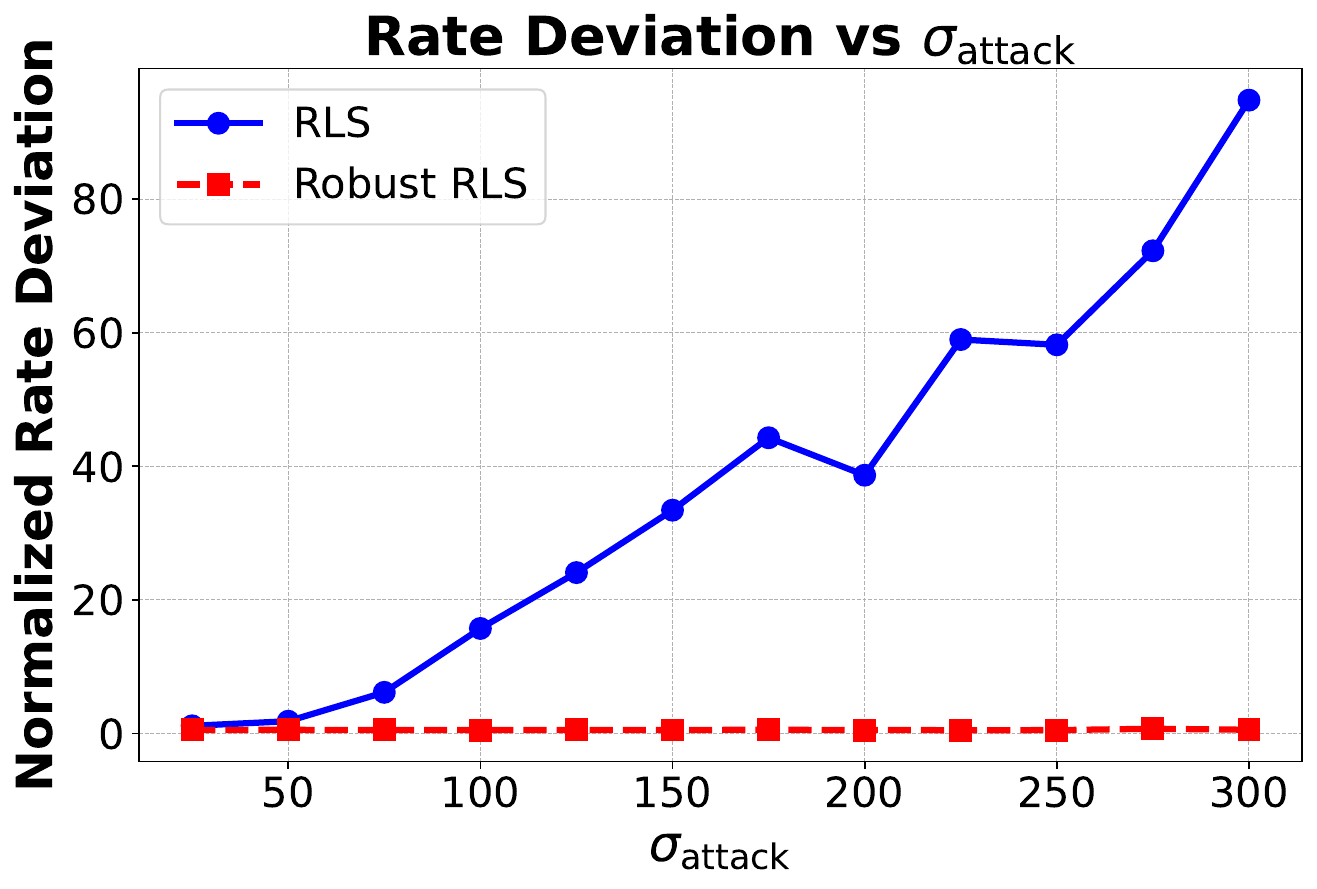}
        \caption{Robust RLS achieves near-zero rate deviation even at extreme (\(\sigma_{\text{attack}}\)).}
        \label{fig:rate-error-vs-sigma}
    \end{subfigure}
    \hspace{0.5cm} % Adjust spacing
    \begin{subfigure}[t]{0.47\textwidth}
        \centering
        \includegraphics[width=\linewidth]{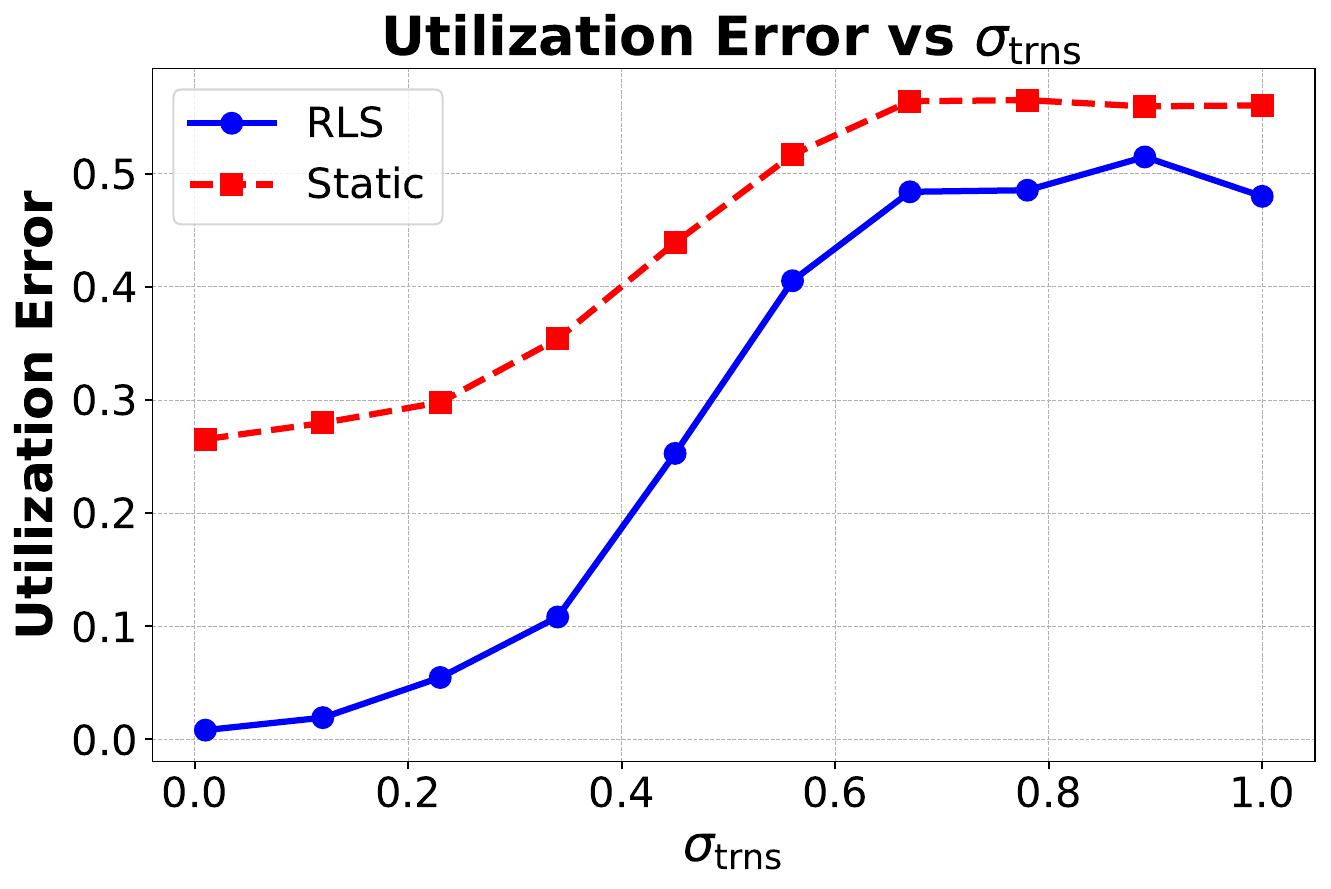}
        \caption{Impact of transition noise on average utilization error for RLS and Aave controllers.}
        \label{fig:util-vs-trns-noise}
    \end{subfigure}

    %\caption{Comparison of the RLS-based interest rate controller with Aave under different conditions.}
\end{figure}

\subsection{Interest rate controller}
To evaluate our interest rate controller, we analyze two demand-supply models: one based on real-world Compound data (Section~\ref{subsec:emperical-validation}) and another using synthetic parameter trajectories. The latter follows a Gaussian random walk, where each parameter (\(\ab, \bb, \al, \bl\)) evolves every \(T\) steps as \(\abn_{t+T} = \ab + \epsilon_t\), with \(\epsilon_t \sim \mathcal{N}(0, \sigmatrns \cdot |\ab|)\). Demand and supply are then generated using Equations~\ref{eq:demand-model} and \ref{eq:supply-model}, incorporating normally distributed noise (\(\sigma_{\text{noise}}\)).

%Notably, these experiments focus solely on the dynamics of demand and supply within the pool after accounting for liquidations and defaults, with the controller remaining agnostic to the risk management module.

\begin{table}[h]
    \centering
    \begin{minipage}[b]{0.48\textwidth}
        \centering
        \resizebox{\textwidth}{!}{ % Adjust table size
        \begin{tabular}{|c|cc|}
        \hline
        \diagbox{Pool}{Utilization MSE} & RLS & Compound \\
        \hline
        DAI  & 0.0085  & 0.0141 \\
        USDC & 0.0009  & 0.0079 \\
        USDT & 0.0001 & 0.0032 \\
        \hline
    \end{tabular}
        }
        %\vspace{1em} % Add vertical space to align with the second table
        \caption{Utilization MSE comparison for RLS-based controller and Compound Static Curve.}
        \label{table:utilization-mse}
    \end{minipage}
    \hfill
    \begin{minipage}[b]{0.48\textwidth}
        \centering
        \resizebox{\textwidth}{!}{ % Adjust table size
        \begin{tabular}{|c|c|c|c|c|}
            \hline
            \sigmatrns & Adaptive FF & \(\ff = 0.85\) & \(\ff = 0.9\) & \(\ff = 0.99\) \\
            \hline
            0.1 & \textbf{51.15} & 54.13 & 52.72 & 51.48 \\
            0.2 & \textbf{51.58} & 54.58 & 53.08 & 56.32 \\
            0.4 & \textbf{56.62} & 58.12 & 58.59 & 107.60 \\
            \hline
        \end{tabular}
        }
        \caption{RMSE of estimated demand function by RLS for different transition noises and forgetting factors.}
        \label{table:variable-ff-rmse-demand}
    \end{minipage}
\end{table}

\noindent\textbf{Utilization error} 
We evaluate our RLS-based controller against Compound’s static interest rate curves tuned for optimal utilization of $0.8$. Using Compound’s data (Feb 2024–Feb 2025), we infer \(\ab, \bb, \al, \bl\) to simulate user behavior. Our controller archives significantly lower utilization mean squared error (MSE) across all pools (Table~\ref{table:utilization-mse}). We use $\ff=0.85$.
To assess robustness, we test synthetic data. As \(\sigmatrns\) increases, inducing greater demand-supply fluctuations, our controller incurs more errors but consistently outperforms the Aave/Compound static approach (Figure~\ref{fig:util-vs-trns-noise}).  
For details, see Appendix~\ref{app:util-opt}.

% We evaluate our RLS-based controller in regulating utilization against Compound’s static interest rate curves. Using Compound’s data (Feb 2024–Feb 2025), we infer \(\ab, \bb, \al, \bl\) to simulate user behavior. With a forgetting factor of 0.85, our controller targets a utilization of 0.8, and its mean squared error (MSE) is compared to Compound’s static curves. As shown in Table~\ref{table:utilization-mse}, the RLS-based controller consistently reduces utilization error across all three pools.
% %
% To assess robustness, we experiment with synthetic data. Figure~\ref{fig:util-vs-trns-noise} shows that as \(\sigmatrns\) increases, causing greater demand-supply fluctuations, our algorithm incurs more errors but consistently outperforms the Aave/Compound static approach.% This highlights its superior adaptability under high transition noise.  
% For details, see Appendix~\ref{app:util-opt}.

\noindent\textbf{Variable Forgetting Factor}  
This experiment evaluates the impact of different forgetting factors on demand estimation under varying parameter evolution rates (\(\ab, \bb\)) and examines whether the adaptive forgetting factor outperforms RLS with a fixed forgetting factor.  
We consider transition noise levels \(\sigmatrns = 0.1, 0.2,\) and \(0.4\). As shown in Table~\ref{table:variable-ff-rmse-demand}, the optimal fixed forgetting factor decreases with increasing noise: \(\ff = 0.99\) performs best at \(\sigmatrns = 0.1\), while \(\ff = 0.9\) and \(\ff = 0.85\) yield lower errors at \(\sigmatrns = 0.2\) and \(0.4\), respectively.  
However, the adaptive forgetting factor consistently achieves the lowest error across all cases.

%
%This superiority stems from its ability to dynamically adjust the forgetting factor: it reduces the factor during rapid parameter changes to prioritize recent data and increases it as estimates stabilize, effectively incorporating historical data to enhance noise cancellation and stability.

% \begin{table}[th]
% \centering
% \begin{tabular}{|c|c|c|c|c|}
% \hline
% \sigmatrns & Adaptive \ff & \(\ff = 0.85\) & \(\ff = 0.9\) & \(\ff = 0.99\) \\
% \hline
% 0.1 & \textbf{51.15} & 54.13 & 52.72 & 51.48 \\
% 0.2 & \textbf{51.58} & 54.58 & 53.08 & 56.32 \\
% 0.4 & \textbf{56.62} & 58.12 & 58.59 & 107.60 \\
% \hline
% \end{tabular}
% \caption{Square root of the mean squared error (RMSE) of the demand function for different transition noise levels and forgetting factor values.}
% \label{table:variable-ff-rmse-demand}
% \end{table}

\noindent\textbf{Robust RLS}  
We assess the robustness of robust RLS against adversarial demand and supply manipulation. The metric of interest is the \textit{normalized rate deviation}, measuring deviation from the optimal rate for \(\uoptimal = 0.7\).  
Two attack strategies are considered. The first is an \textit{intermittent adversary}, which randomly activates with probability 0.1, increasing the demand and supply noise variance up to \(\sigma_{\text{attack}}\) (20\%–300\% of demand or supply). As shown in Figure~\ref{fig:rate-error-vs-sigma}, plain RLS exhibits sharp deviation growth, whereas robust RLS maintains near-zero deviation.  
The second strategy simulates a \textit{persistent adversarial borrower} who inflates \(\ab\) by \(\gamma_{\text{adv}}\) to exaggerate demand sensitivity, tricking the protocol into setting lower rates. With \(T_{\text{attack}} = 100\) and \(\gamma_{\text{adv}} \in \{2, 5, 10, 20\}\), plain RLS suffers a normalized rate deviation of up to 3.13, meaning the rate reaches 3.13 times its optimal value. Robust RLS keeps it below 0.5, demonstrating resilience.

%

% \begin{wrapfigure}{r}{0.5\textwidth} % 'r' for right, and 0.4\textwidth for figure width
%     \centering
    
% \end{wrapfigure}

\subsection{Risk controller}
%\noindent\textbf{Controlling liquidation} 
Figure \ref{fig:risk-controller} shows our risk controller adjusting the collateral factor to liquidation threshold ratio as Ethereum's volatility changes. The goal is to maintain an average liquidation rate of 1\% (orange) or 0.1\% (blue), with a fixed liquidation threshold of $\liqthrshN_t = 0.9$. The controller lowers the collateral factor during high volatility to prompt borrowers to add collateral and avoid liquidation.

%Figure \ref{fig:compare-util-actual-sim-dai} illustrates how the RLS-based controller operates on user behavior parameters learned from the actual data of Aave's DAI pool. The blue line represents the realized utilization in the protocol, while the green line indicates the desired utilization set by the protocol. It is evident that Aave's protocol struggles to effectively stabilize at the desired utilization, whereas the RLS-based algorithm demonstrates superior stability in achieving this goal.

%% file: conclusion.tex
\section{Conclusion}

\noindent\textbf{Adaptivity and robustness}  
In this work, we introduced a model of non-stationary borrower and lender behavior that aligns with empirical data, along with an interest rate controller designed to maintain stability under dynamic market conditions. We also characterized the protocol's responsiveness to market changes and analyzed the limits of this adaptivity when facing adversarial manipulation. We suggested solutions to reduce adversarial manipulation.

\noindent\textbf{Limitations and future work}  
Our model has three key limitations, suggesting directions for future work. First, it assumes lending does not impact collateral prices, which may not hold for low-liquidity assets. Second, it considers only a single lent-collateral pair, whereas a broader model would account for multiple assets with correlated price and volatility movements. Third, we focus on Compound-style pools, where collateral remains idle, unlike Aave’s cross-asset model, which generates yield but adds complexity to the risk analysis of the protocol. Due to these risk management challenges, Compound opts for a simpler design, which we also adopt in our analysis \cite{compoundV3}.
% Our model has three key limitations, leaving room for future research. First, we assume that lending does not affect the collateral’s price, which may not hold true for low-liquidity assets. Second, we focused on a single pair of lent and collateral assets. A more comprehensive model would adjust interest rates for a broader range of assets, considering interactions between multiple borrowed and collateralized assets, with potentially correlated price and volatility movements. Third, we focus on Compound-style borrow-lending pools, where borrowers' collateral remains idle and cannot be lent out to earn interest. In contrast, platforms like Aave use a cross-asset model, allowing collateral to generate yield. While this approach offers additional returns, it introduces complex risk management challenges \cite{compoundV3}. As a result, protocols like Compound have opted for a simpler model, which we also adopt in our analysis. 
%
Finally, leveraging neural networks and deep reinforcement learning could enhance user modeling and protocol design. However, their computational inefficiency hinders on-chain use, even with rollups, and their lack of interpretability challenges financial transparency. Developing efficient, interpretable algorithms remains an open problem.

% \commentMahsa{we can get rid of this part >>} Thirdly, \prettyref{sec:adversarial} demonstrates the vulnerability of a fully automated adaptive approach to adversaries, calling for manual/governance guardrails while bringing automation to DeFi. Robustness to adversaries may also be increased by establishing complementary interest rate derivatives markets for each lending pool. Several protocols exist already to provide this service, but systematically valuing their contracts is a problem that has not been solved yet. Future work can focus on fleshing out a how on-chain derivatives like yield tokens and interest rate swaps should be priced. This would equip honest lenders and borrowers with a bare minimum criteria to base their investment decisions on, and make the risks associated with lending more transparent.

\section{Acknowledgment}
This work was supported by the National Science Foundation via grants CCF-1705007, CNS-2325477, the Army Research Office via grant W911NF2310147, a grant from C3.AI and a gift from XinFin Private Limited.

%% file: appendix.tex
\section{Details for our algorithms}
In this section, we detail the algorithms we use in \prettyref{sec:interestrate-controller}.
\input{algorithms/estimator}

\input{algorithms/optimizer}

\section{Supplementary explanations about the model and the results }
\subsection{Related work}\label{app:related}
\noindent\textbf{Collateralized lending in traditional finance}  
Collateralized lending has been extensively studied in traditional finance, where the main challenge is information asymmetry between borrowers and lenders, leading to issues like moral hazard and credit rationing \cite{chan1985asymmetric, stiglitz1981credit}. In contrast, DeFi platforms benefit from shared access to crypto price history, allowing for more transparent, algorithmic risk assessment. The key challenge in DeFi borrow-lending is determining a fair, adaptive interest rate in a rapidly changing, competitive environment, unlike traditional finance where rates change infrequently and competition is limited.

\noindent\textbf{Collateralized lending in decentralized finance and adaptive protocols}  
Various models have explored lender and borrower behavior in DeFi, focusing on interest rate equilibria and protocol efficiency. Some models propose equilibrium-based rates but overlook external markets and default risk strategies \cite{cohen23}. Others consider external market dynamics but lack focus on long-term decisions and liquidation risks \cite{rivera2023equilibrium}. Nash equilibrium studies highlight how protocol-driven prices can cause oscillations and require contract adjustments \cite{chiu2022fragility}. Empirical data has helped refine models of user behavior \cite{aave_gauntlet,compound_gauntlet,compoundEmpirical}, and adversarial risks to DeFi protocols have been examined, exposing vulnerabilities \cite{chitra2023attacks,cohen2023paradox,carre2023security}. Recent work on adaptive financial mechanisms addresses impermanent loss and arbitrage in market makers \cite{AIeconomist, goyal2023finding, milionis2023automated, nadkarni2023zeroswap,nadkarni2024zeroswap}. Strategic behavior exploiting static rate curves has also been identified \cite{yaish2023suboptimality}. Protocols like Morpho \cite{morpho} and Ajna \cite{ajna} support adaptive rate discovery but require continuous monitoring. Our work is closest to \cite{bastankhah2024thinking} which proposes a two-timescale adaptive lending protocol. However, it uses a simplistic model of user behaviour and only considers shortsighted adversaries. While \cite{bastankhah2024thinking} focuses on a specific demand-supply dynamic yielding a single equilibrium rate, we use a simpler linear demand and supply model with varying parameters, supported by empirical data. This allows for a range of possible equilibrium rates, enabling the protocol to select the one that best achieves the desired utilization.\\
\subsection{Explicit definition of various objectives}\label{app:exp_obj}
\subsubsection{Demand-supply balancing}\label{subsec:demand-supply-balancing}

In decentralized finance (DeFi) platforms, maintaining an optimal balance between supply and demand is crucial, and a key metric for balance is the platform’s proximity to the target utilization rate, \(\uoptimal\) \cite{aave_gauntlet,compound_gauntlet}. Low utilization indicates that deposited supply is not being efficiently utilized, while high utilization negatively impacts user experience, as it restricts lenders from withdrawing their funds, effectively locking them up.

To assess the effectiveness of a protocol \(\protocol\) in achieving optimal utilization, we define the optimal interest rate and measure the deviation of the protocol's interest rate from this target. Specifically, we consider a pool where demand and supply are modeled as noisy versions of \(f(r; \colborrowers)\) and \(g(\effrate;\collenders)\), respectively with a Gaussian noise with standard deviation $\nu$. At time \(t_0\), the pool variables \(\colborrowers\) and \(\collenders\) shift from their initial values to \(\mathcal{B}_0\) and \(\mathcal{L}_0\), after which they remain constant. This allows the system to stabilize at a certain utilization \(U\) and interest rate \(r\). The optimal interest rate \(r^*_{\text{utl}}\) is defined as:
\begin{equation}\label{eq:optimal-rate-for-u}
    r^*_{\text{utl}} \coloneqq \arg\min_{r} \left|U - \uoptimal\right| \quad \text{subject to} \quad U = \frac{f(r; \mathcal{B}_0)}{g(\effrate; \mathcal{L}_0)},\; \effrate \coloneqq U\cdot r
\end{equation}
\noindent We propose a metric called \emph{\rateDeviation} denoted by $\rateDeviationN$ to assess the \protocol's performance in maintaining $\uoptimal$:
\begin{equation}\label{eq:utilization-rate-deviation}
    \rateDeviationN \coloneqq \mathbb{E}\left[\left|r_t -  r^*_{\text{utl}}\right|\right]
\end{equation}
\noindent The expectation is taken over the protocol's randomness, as well as the noise present in the users' demand and supply. \footnote{A more pragmatic version of demand-supply balancing that maximizes the revenue of the protocol is formally defined in \prettyref{app:alt_util_objective} and \prettyref{app:revenue_max}}
\subsubsection{Adversarial robustness} 
Strategic borrowers and lenders seek to manipulate the protocol's interest rate to their advantage by exploiting the structure of the interest rate controller. They may simulate the protocol’s responses to various demand and supply scenarios, selecting the one that maximizes their own utility rather than acting truthfully. We aim to quantify how much influence these strategic users can exert on the interest rate.
Specifically, we consider a lending pool governed by the protocol \(\protocol\) with a set of truthful users characterized by fixed demand and supply curves \(f(r;\colborrowers)\) and \(g(\effrate;\collenders)\); Plus one strategic lender, \advl, who has an external interest rate \(r^l\) and a deposit of \(\delta_l \times \big(\max_{x} g(x;\collenders)\big)\), and one strategic borrower, \advb, with an external rate \(r^b\) and demand of \(\delta_b  \times\big(\max_{x} f(x;\colborrowers)\big)\).

Let \( r_{\text{truthful}} \) and \( r_{\text{strategic}} \), respectively denote the steady-state interest rates set by \protocol if \advl and \advb behave truthfully versus strategically. We introduce a new metric, \emph{\advImpact}, to formally quantify the impact of strategic users on interest rate manipulation:
\begin{equation}
    \advImpactN \coloneqq  \sup_{r^b,r^l}\E{\left|r_{\text{truthful}} - r_{\text{strategic}} \right|}
\end{equation}

\subsubsection{Liquidation and default risk control}
In a peer-to-pool-to-peer lending platform, as discussed in Section \ref{subsec:market-actors}, the main risks are defaults for lenders and liquidations for borrowers. To mitigate these risks, the protocol must dynamically adjust parameters like the collateral factor \(\colfacN_t\), liquidation threshold \(\liqthrshN_t\), and liquidation incentive \(\liqIncntive_t\) in response to price volatility.

Pool defaults occur when a borrower's debt exceeds the value of their collateral due to a price drop, making the pool unable to recover the full loan amount. Liquidation occurs when a borrower, maintaining the maximum loan-to-value ratio, faces a price drop that forces the protocol to liquidate part of their collateral to bring the ratio back below the liquidation threshold. These risks are formally defined in \prettyref{app:risk-metrics}, where the exact conditions for pool default and liquidation are detailed.

The protocol's effectiveness in managing these risks can be measured using metrics like the expected value or the 95th percentile of pool defaults and liquidations, computed with respect to the price distribution.
\subsection{Alternate objective for demand/supply balancing}\label{app:alt_util_objective}

Traditionally, balance in Defi lending has been measured by how closely the utilization rate, \(\ut\), approaches its optimal level, \(\uoptimal\). However, this approach can be limiting, particularly in situations where supply is high and demand is low. In such cases, maintaining utilization at a preset level requires setting a very low interest rate, which discourages lenders and may ultimately undermine the system's revenue.
To address these limitations, we broaden the definition of balance beyond utilization alone. We define an optimal pool state as one characterized by specific values of \(\lt\), \(\bt\), and \(\ut\), and measure how closely the protocol approaches the interest rate that best satisfies this state. 
One example of this objective is maximizing the combined demand and supply while keeping utilization below a safe threshold \(U_{\text{max}}\), thereby enhancing both revenue and user experience. The optimal interest rate, according to this new metric, is given by:
\begin{equation}\label{eq:balance-metric-r}
    r^*_{\text{rev}} \coloneqq \arg\max_{r} \left[f(r; \theta) + g(rU; \omega)\right] \quad \text{subject to} \; U < U_{\text{max}} \; \text{where} \; U = \frac{f(r; \theta)}{g(rU; \omega)}
\end{equation}
We can define an interest rate deviation metric for this objective, akin to the utility-optimizing case in Equation \ref{eq:utilization-rate-deviation}. 

\subsection{Optimizing interest rate to maximize revenue} \label{app:revenue_max}
% We can select an interest rate with various objectives in mind, such as maximizing the sum of demand and supply, which reflects the protocol's revenue. To achieve this, we first need to formulate an optimization problem that determines the optimal interest rate, \(\rt\). Subsequently, we will develop an algorithm, similar to Algorithm \ref{alg:optimizer}, that selects the interest rate using a revenue maximization formula.
Maximizing revenue can be also considered as an objective to set the interest rate, in order to do that it turns out that we need a parameter estimator exactly the same as Algorithm \ref{alg:rls-forgetting-factor} and an optimizer similar to Algorithm \ref{alg:optimizer} but the formula for the optimal rate based on the estimated parameters is given in the following theorem:

\begin{theorem}\label{theorem:rev-max}
Given the user behavior models described in Equations \ref{eq:supply-model} and \ref{eq:demand-model}, the revenue maximization problem, as described in Equation \ref{eq:balance-metric-r}, can be formulated as follows:
\begin{align}
    &\max_{r} \; \big(-\abn \,r  + \aln \, r \,U \big)\\ &\text{Subject to}\quad U = \,\frac{\bln + \sqrt{(\bln)^2 - 4\aln \,r (\abn \,r - \bbn)}}{2\aln \,r}\quad  
 \text{and }\quad U < \umax\notag
\end{align}
And its solution is:
\begin{align}\label{eq:rev-optimal-rate}
    r_{\text{rev}}^* = 
    \begin{cases} 
    \frac{\bbn + \bln \umax}{\abn + \aln (\umax)^2} & \text{if  } \frac{\abn\,\bln (1 + \sqrt{1 - 4\, \abn \, \aln})}{\aln \left( \bbn + \sqrt{(\bbn)^2 + 4\, (\abn\,\bln)^2} \right)} \geq \umax \\
     \frac{\bbn + \sqrt{(\bbn)^2 + 4\,(\abn\,\bln)^2}}{2\, \abn}
& \text{O.W. }  
    \end{cases}
\end{align}

\end{theorem}
Informally speaking, replacing the expected optimal rate at line \ref{alg-line:optimal-rate} of Algorithm \ref{alg:optimizer} with Equation \ref{eq:rev-optimal-rate} will result in an RLS-based revenue-maximizing rate controller, offering similar convergence guarantees as those outlined in Theorem \ref{theorem:rate-deviation-RLS}.

\subsection{Formal abstraction of the learning-based controller }
\label{app:lc_auction}

Formally, \learningController is implemented as follows:
\begin{enumerate}
    \item In the exploration phase, each borrower \(i\) privately reports a demand quantity \(\hat{B}_t(i)\) along with the maximum interest rate they are willing to pay as their bid \(\hat{r}^b_t(i)\). Similarly, each lender \(i\) privately reports their supply quantity \(\hat{L}_t(i)\) and their bid \(\hat{r}^l_t(i)\).
    
    \item \learningController determines the demand and supply curves \(\hat{f}(r;\theta)\) and \(\hat{g}(r\, U;\omega)\) from the reported values as follows:
    \[
    \hat{f}(r;\theta) = \sum_{i} \hat{B}_t(i) \cdot \indicator{\left(\hat{r}^b_t(i) \geq r\right)},
    \]
    \[
    \hat{g}(r\, U;\omega) = \sum_{i} \hat{L}_t(i) \cdot \indicator{\left(\hat{r}^l_t(i) \leq r\, U\right)},
    \]

    \item \learningController sets the interest rate as:
    \(
    \rt = \arg\min_{r} \left|U - \uoptimal\right| \, \text{subject to} \, U = \frac{\hat{f}(r; \theta)}{\hat{g}(rU; \omega)}
    \)
    
    \item The utility of borrower \(i\) at time \(t\) is defined by:
    \[
    \text{Utility}^B_{i,t} (\mathbf{\hat{B}}_t, \mathbf{\hat{L}}_t, \mathbf{\hat{r}}^b_t, \mathbf{\hat{r}}^l_t) \coloneqq
    \begin{cases} 
        -\hat{B}_t(i)\,r_t - \left(B_t(i) - \hat{B}_t(i)\right) \,r^b_t(i), & \text{if } r_t \leq \hat{r}^b_t(i),  \\
        -{B}_t(i)\,r^b_t(i), & \text{otherwise}.
    \end{cases}
    \]
    Here, the vectors \(\mathbf{\hat{B}}_t\), \(\mathbf{\hat{L}}_t\), \(\mathbf{\hat{r}}^b_t\), and \(\mathbf{\hat{r}}^l_t\) represent the bids and quantities for all participants. And $r^b_t(i)$ is the true external rate of borrower $i$. Similarly, the utility of lender \(i\) at time \(t\) is given by:
    \[
    \text{Utility}^L_{i,t} (\mathbf{\hat{B}}_t, \mathbf{\hat{L}}_t, \mathbf{\hat{r}}^b_t, \mathbf{\hat{r}}^l_t) \coloneqq
    \begin{cases} 
        \hat{L}_t(i)\,r_t \,U + \left(L_t(i) - \hat{L}_t(i)\right)\, r^l_t(i), & \text{if } r_t \geq \hat{r}^l_t(i), \\
        L_t(i)\,r^l_t(i), & \text{otherwise}.
    \end{cases}
    \]
\end{enumerate}

\subsection{Utility Functions for Strategic Withholding}\label{app:utility-static}

In this section, we provide the detailed utility functions for strategic borrowers and lenders who engage in the \emph{strategic withholding} strategy.

The utility of a strategic borrower \(\advb\), with external rate \(r^b\), who borrows an amount \(\hat{B} \leq \delta_b B\) from the pool while borrowing the rest from an external market, is given by:
\[
\text{Utility}_{\advb}(\hat{B}) \coloneqq -\hat{B} \cdot \phi\left(\frac{B(1-\delta_b) + \hat{B}}{L}\right) - \big(B \cdot \delta_b - \hat{B}\big) \, r^b,
\]
where \(\phi(.)\) denotes the baseline interest rate curve (see Equation \ref{eq:baseline-controller}).

Similarly, the utility of a strategic lender \(\advl\), with external rate \(r^l\), who deposits \(\hat{L} \leq \delta_l L\) in the pool and allocates the remainder to an external market, is given by:
\[
\text{Utility}_{\advl}(\hat{L}) \coloneqq \hat{L} \cdot \phi\left(\frac{B}{L(1-\delta_l) + \hat{L}}\right) \cdot \frac{B}{L(1-\delta_l) + \hat{L}} + \big(L \cdot \delta_l - \hat{L}\big) \, r^l.
\]
These utility functions are used to determine the worst-case adversarial impact, where strategic users maximize their utilities by adopting the strategic withholding strategy.

\subsection{Risk Metrics for Default and Liquidation}\label{app:risk-metrics}

In this section, we provide the formal definitions for pool default and liquidation.

The pool default between timeslot \(t\) and \(t+1\), due to a price change from \(\pt\) to \(\pS{t+1}\), is defined as:
\begin{equation}\label{eq:userDefault}
    \poolDefault{t}(\pS{t+1}) \coloneqq \sum_{i \in \text{Borrowers}}\max\left\{0, \bS{t}(i) - \cS{t}(i) \cdot \pS{t+1}\right\}
\end{equation}
This equation captures the total amount by which the borrowers' debt exceeds their collateral value, resulting in a default.

For liquidation, consider a borrower \(i\) who maintains the maximum loan-to-value ratio allowed by \protocol at time \(t\). If the price drops at time \(t+1\), the borrower may need to undergo a liquidation to ensure the loan-to-value ratio stays below the liquidation threshold. The minimum required liquidation amount, \(\userLiq{t}{i}(\pS{t+1})\), is given by:
\begin{align*}
    \userLiq{t}{i}(\pS{t+1}) \coloneqq \min \left\{ x \ \middle|\  \frac{\bt(i) - x}{\ct(i) \,\pS{t+1} - x \,(1+\liqIncntive_t)} \leq \liqthrshN_t \right\}
\end{align*}
where \(\bS{t+1}(i) = \bt(i) - x\) is the remaining debt after liquidation, and \(\cS{t+1}(i) = \ct(i) - x\,(1 + \liqIncntive_t)\) is the remaining collateral after the liquidator receives their reward.

These risk metrics allow us to quantify the potential impact of price volatility on the protocol's stability.

\subsection{M-estimate function for robust RLS}\label{app:m-estimate-function}
We adopt the M-estimate function proposed by \cite{zou2000recursive}:
\begin{equation}
\phi(e) \triangleq
\begin{cases} 
\frac{e^2}{2}, & 0 < |e| < \xi, \\[8pt]
\xi|e| - \frac{\xi^2}{2}, & \xi \leq |e| < \Delta_1, \\[8pt]
\frac{\xi}{2}(\Delta_2 + \Delta_1) - \frac{\xi^2}{2} + \frac{\xi(|e| - \Delta_2)^2}{\Delta_1 - \Delta_2}, & \Delta_1 \leq |e| < \Delta_2, \\[8pt]
\frac{\xi}{2}(\Delta_2 + \Delta_1) - \frac{\xi^2}{2}, & \Delta_2 \leq |e|.
\end{cases}
\end{equation}

Following \cite{zou2000recursive}, the thresholds \(\xi\), \(\Delta_1\), and \(\Delta_2\) are determined using confidence levels of \( 0.05 0.025\), and \( 0.01\). Under the assumption of Gaussian noise, these thresholds are computed as \(\xi = 1.96\hat{\sigma}(t)\), \(\Delta_1 = 2.24\hat{\sigma}(t)\), and \(\Delta_2 = 2.576\hat{\sigma}(t)\), where \(\hat{\sigma}(t)\) is the estimated standard deviation of the noise.

\section{Complimentary Material of Evaluation}
\subsection{Parameter estimation of Compound demand and supply curves}\label{app:eval-estimated-params}
We analyze data from the three major lending pools—DAI, USDC, and USDT—on Compound V2 over the period from February 9, 2024, to February 9, 2025. Compound V2 uses different interest rates for borrowing and supplying assets. To model the relationship between past interest rates and subsequent demand and supply, we estimate the functional dependence between the interest rate at time \( t - \Delta \) and the corresponding demand and supply at time \( t \).

A key challenge is determining the appropriate time lag \( \Delta \), as the response time of users to interest rate changes is not known a priori. To identify a reasonable lag, we evaluate different values of \( \Delta \) ranging from 1 to 30 time slots (each time slot corresponding to a three-hour interval). For each candidate \( \Delta \), we apply Recursive Least Squares (RLS) with a forgetting factor of \( \ff = 0.95 \) and estimate future demand and supply. The optimal lag is selected based on the value that minimizes the mean squared error (MSE) of the RLS predictions.

The results indicate that users of the DAI pool exhibit slower reactions to interest rate changes compared to those in USDC and USDT pools. Specifically, the optimal lag for the DAI demand and supply curves is found to be \( \Delta_{\text{DAI, demand}} = 26 \) and \( \Delta_{\text{DAI, supply}} = 16 \), whereas for USDC and USDT, the optimal lag is \( \Delta = 1 \). 
Using these optimal shift values, we apply RLS to estimate the evolution of the demand and supply parameters (\(\ab, \bb, \al, \bl\)) over time for each pool. %The resulting estimated parameters for the three pools are presented in Figures~\ref{}. 

% \begin{figure}[ht]
% \centering
% \includegraphics[width=1\textwidth]{images/cUSDC_parameter_evolution.pdf}
% \caption{Estimated parameters for the demand curve. The top plot shows the estimated $\ab, \bb$ parameters over time for USDC pool, and the bottom plot shows the estimated $\al, \bl$ parameters over time for USDC. RLS algorithm was used with forgetting factor $0.95$.}
% \label{fig:demand_params}
% \end{figure}

% \begin{figure}[ht]
% \centering
% \includegraphics[width=1\textwidth]{images/cUSDT_parameter_evolution.pdf}
% \caption{Estimated parameters for the supply curve. The top plot shows the estimated $\ab,\bb$ parameters over time for the USDT pool, and the bottom plot shows the estimated $\al,\bl$ parameter over time for USDT. RLS algorithm was used with forgetting factor $0.95$.}
% \label{fig:supply_params}
% \end{figure}

% \begin{figure}[ht]
% \centering
% \includegraphics[width=1\textwidth]{images/cDAI_parameter_evolution.pdf}
% \caption{Estimated parameters for the supply curve. The top plot shows the estimated $\ab, \bb$ parameters over time for DAI, and the bottom plot shows the estimated $\al, \bl$ parameters over time for DAI. RLS algorithm was used with forgetting factor $0.95$.}
% \label{fig:supply_params}
% \end{figure}

\subsection{Utilization optimization}\label{app:util-opt} 
%\commentMahsa{compelete this for the details of the experiment of the inferred demand and supply functions} Refer to Figure \ref{fig:compare-util-actual-sim-dai}.

We now provide the detailed specifications of the experiment used to generate Figure~\ref{fig:util-vs-trns-noise}. The synthetic parameters \(\ab, \bb, \al, \bl\) evolve via a Gaussian random walk with an update interval of \(T = 25\), following:  
\[
\abn_{t+T} = \ab + \epsilon_t, \quad \epsilon_t \sim \mathcal{N}\left(0, \sigmatrns \cdot |\ab|\right),
\]
with similar updates for \(\bb, \al, \bl\). The noise variance is set to \(\sigma_{\text{noise}} = 1\), while \(\sigma_{\text{trans}}\) varies from 0.1 to 1 (1\% to 100\%).  
Simulations run for 1000 time steps, averaging results over 50 independent runs. The initial parameter values are:  
\[
\abn_0 = -10, \quad \abn_0 = 5000, \quad \aln_0 = 500, \quad \bln_0 = -50.
\]
Desired utilization is $\uoptimal = 0.7$ and forgetting factor is $0.8$.

\subsection{Robust RLS experiment}
We evaluate the robustness of the robust RLS algorithm against adversarial manipulation of demand and supply using two distinct strategies. The metric of interest is the \textit{normalized rate deviation}, defined as the deviation of the interest rate from the optimal rate that achieves the target utilization \(\uoptimal = 0.7\).
The first strategy is an intermittent adversary, which randomly activates in every timeslot with a probability of 0.1 to enforce large noise on demand and supply. When inactive, the noise variance is \(\sigmanoise = 10\), and when active, it increases to \(\sigmanoise = |\bt| \cdot \sigma_{\text{attack}}\) for demand and \(\sigmanoise = |\lt| \cdot \sigma_{\text{attack}}\) for supply, where \(\sigma_{\text{attack}}\) varies from 0.2 to 3 (20\% to 300\% of demand or supply). As shown in Figure~\ref{fig:rate-error-vs-sigma}, the normalized rate deviation for plain RLS increases sharply with \(\sigma_{\text{attack}}\), while the robust RLS algorithm maintains near-zero deviation, demonstrating resilience to such attacks.
The second strategy simulates a persistent and more realistic adversarial borrower, which activates with probability 0.01 and stays active for \(T_{\text{attack}}\) steps. The adversary inflates demand using \(\ab r_t \cdot \gamma_{\text{adv}} + \bb\), where \(\gamma_{\text{adv}} > 1\) exaggerates sensitivity to interest rates, tricking the protocol into lowering rates. Stress tests with \(T_{\text{attack}} = 100\) and \(\gamma_{\text{adv}} \in \{2, 5, 10, 20\}\) show that while plain RLS suffers increasing normalized rate deviation (up to 3.13 for \(\gamma_{\text{adv}} = 20\)), robust RLS keeps it below 0.5, even under extreme attacks

These experiments were conducted using synthetic demand and supply functions, following the experimental setup described in Appendix~\ref{app:util-opt}.

\subsection{Forgetting factor and estimation error}
\begin{figure}[t]
    \centering
    \includegraphics[width=\linewidth]{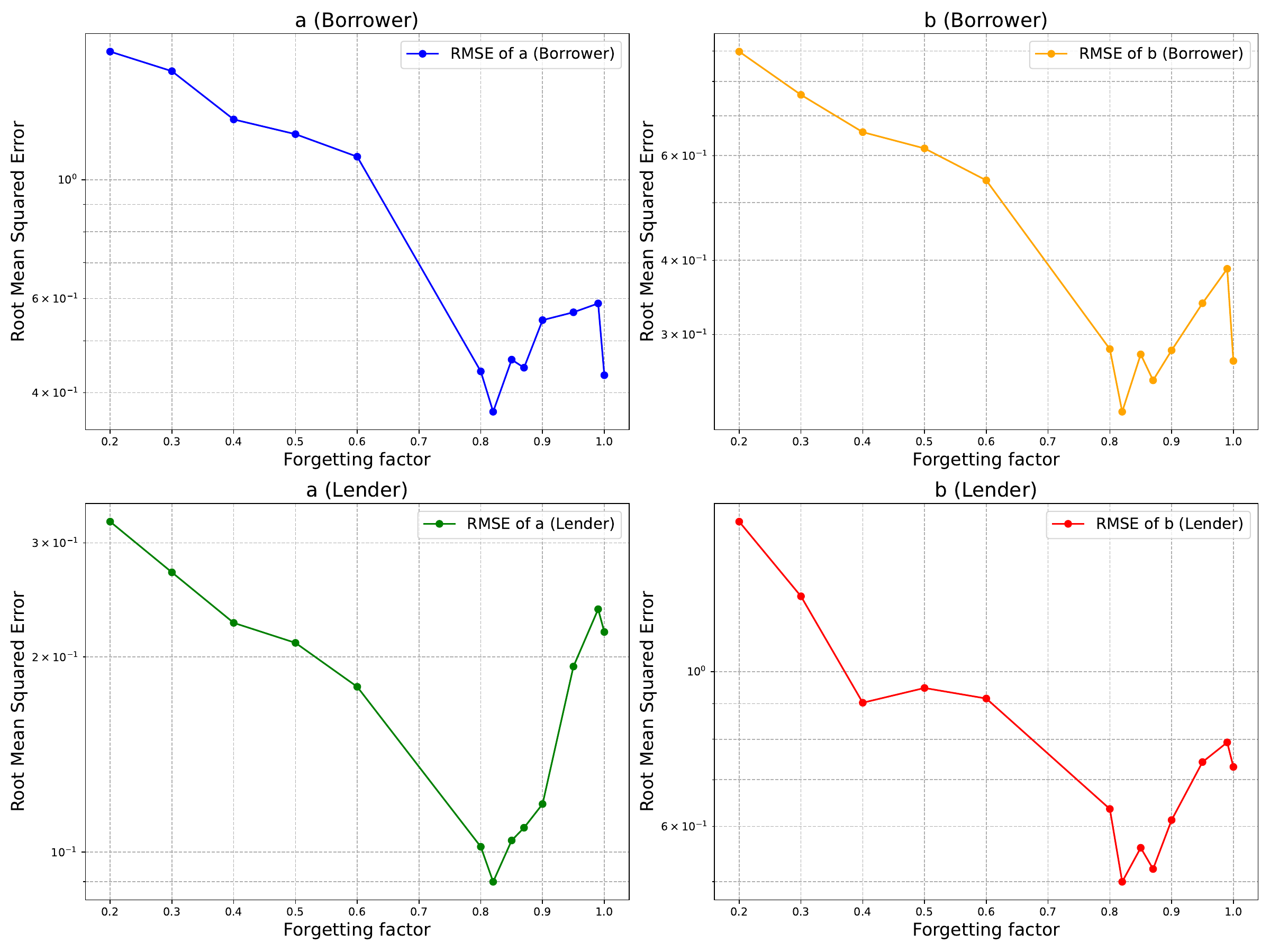}
    \caption{Impact of forgetting factor on the mean square error of the estimated parameters by the RLS-based algorithms.}
    \label{fig:forgettingfactor}
\end{figure}

The forgetting factor governs the trade-off between adaptivity and precision; therefore, depending on the dynamics of the demand and supply parameters, different forgetting factors should be used. Figure \ref{fig:forgettingfactor} illustrates the mean squared error (MSE) of the parameters $\ab, \bb, \al, \bl$ estimated by the RLS-based algorithm. The parameters were generated using a random walk with Gaussian noise, with a frequency of change every $T=50$ steps and $\sigmatrns = 0.1$. We run the experiment for $1000$ timeslots and average the results across $50$ independent runs. In this experiment setup, the optimal forgetting factor, which minimizes the overall MSE across all parameters, is approximately $\ff = 0.82$.

\subsection{Revenue maximizing}

\begin{figure}[t]
\centering
\includegraphics[width=0.7\textwidth]{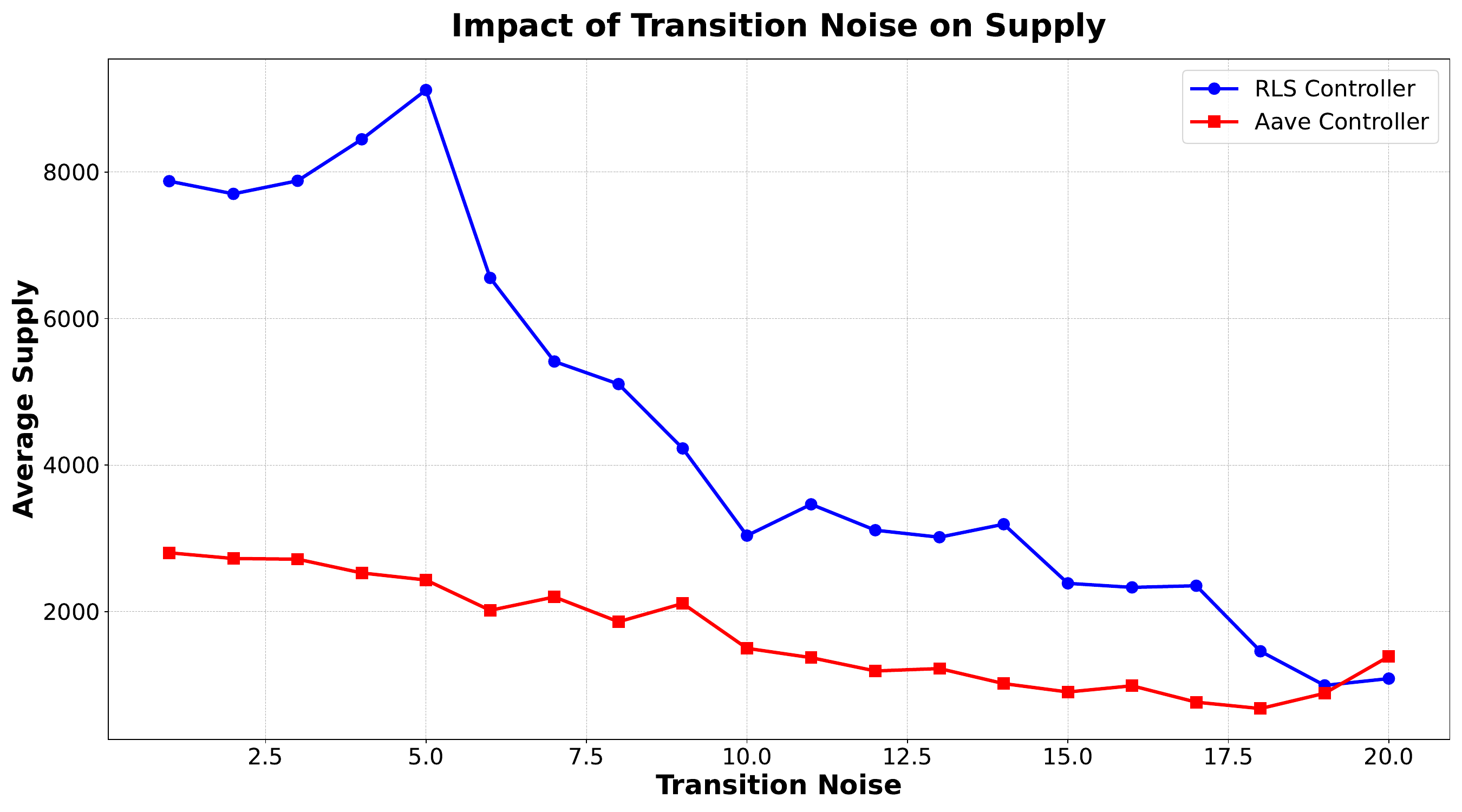}
\caption{Comparison of supply, used as a proxy for revenue, between the RLS-based controller and Aave's static curves. The parameters evolve according to a random walk with Gaussian noise. The x-axis represents the relative standard deviation of the noise in percentage. The fixed factor is set to $\ff=0.8$.}
\label{fig:supply-vs-trns-noise}
\end{figure}

We run our RLS-based controller with the objective of maximizing the supply, which serves as a proxy for the pool's revenue. The user behavior parameters are modeled as random walk stochastic processes with Gaussian noise. In figure \ref{fig:supply-vs-trns-noise} we plot the supply against the std of the transition noise in the process. As the noise increases, it becomes more difficult for the RLS-based algorithm to adapt, resulting in a performance decline. When the relative standard deviation of the noise reaches 20\%---meaning the parameters change with severe 20\% noise at every time slot---the performance of the RLS-based algorithm drops to a level comparable to Aave-style algorithms.

% \begin{figure}[!t]
% \centering
% \includegraphics[width=\textwidth]{images/reality-vs-rlsController-DAI.pdf}
% \caption{Comparison of utilization between RLS-based controller and Aave's static curves, with user supply and demand curves learned from real Aave DAI pool data, $\ff=0.8$.}
% \label{fig:compare-util-actual-sim-dai}
% \end{figure}

\subsection{Risk controller}

\begin{figure}[th!]
    \centering
    \includegraphics[width=\linewidth]{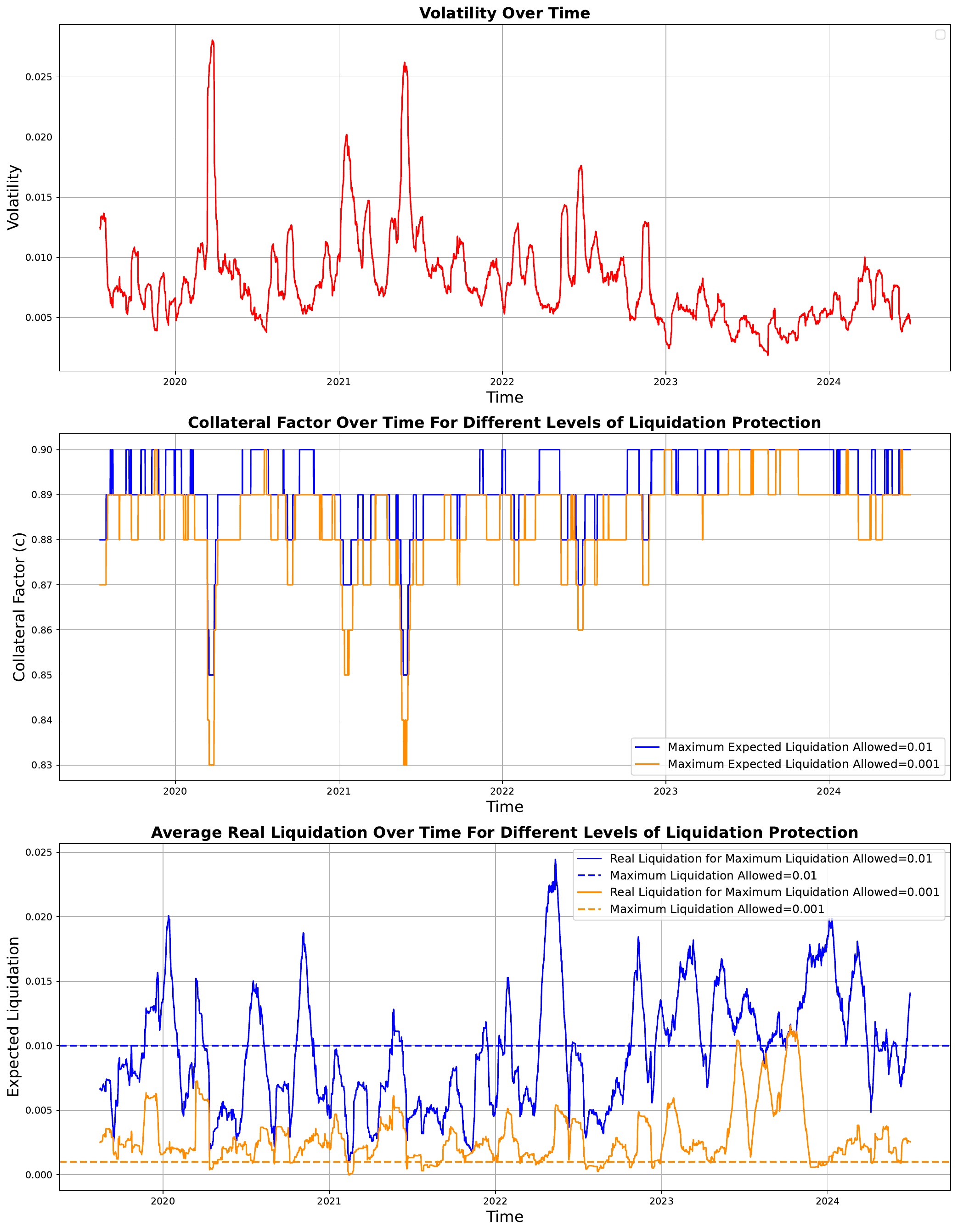}
    \caption{ETH volatility over time, with the controller dynamically adjusting the collateral factor. The bottom plot compares the actual liquidations resulting from this adjusted collateral factor with the target expected liquidations, $\liqthrshN_t = 0.9$.}
    \label{fig:risk-controller}
\end{figure}

\section{Proofs}

\subsection{Proof of theorem \ref{theorem:rate-deviation-RLS}}
\begin{proof}

Our proof is applying a slightly modified version of the proof provided in \cite{canetti1989convergence}. The condition required for convergence of the RLS algorithm is that the input vector of the linear system should be persistently excited. In other words if we want to estimate  $\mathbf{\theta}_t$, and the linear system is $y_t = \mathbf{x}_t^T \mathbf{\theta}_t + e_t $ then $\mathbf{x}_t$ should hold in:
\begin{equation*}
    mI \leq \sum_{i=t}^{t+N} \mathbf{x}_i\mathbf{x}_i^\top \leq MI, \forall t > 0
\end{equation*}
for some positive $m,M,N$. We first prove the following claim:

\textbf{Claim:} Let \( x_i = \begin{bmatrix} 1 \\ r_i \end{bmatrix} \) for \( i = 1, 2, \dots, N \), where \( r_i \in \mathbb{R} \), and suppose at least two of the \( r_i \) are distinct. Then, the matrix
\[
S = \sum_{i=t}^{t+N} x_i x_i^\top
\]
is positive definite, satisfying the persistent excitation condition.
In order to show the claim holds, we compute the sum matrix \( S \):

\[
S = \sum_{i = t}^{t+N} x_i x_i^\top = \sum_{i = t}^{N} \begin{bmatrix} 1 \\ r_i \end{bmatrix} \begin{bmatrix} 1 & r_i \end{bmatrix} = \sum_{i = t}^{N} \begin{bmatrix} 1 & r_i \\ r_i & r_i^2 \end{bmatrix} = \begin{bmatrix} N & \sum_{i = t}^{N} r_i \\ \sum_{i = t}^{N} r_i & \sum_{i = t}^{N} r_i^2 \end{bmatrix},
\]
We compute \( \det(S - \lambda I) \) to find the eigenvalues:
\[
(N - \lambda)(c - \lambda) - \left( \sum_{i = t}^{N} r_i \right)^2 = 0.
\]
Let:
\(
b \coloneqq \sum_{i = t}^{N} r_i, \quad c \coloneqq \sum_{i = t}^{N} r_i^2.
\)

\[
\lambda_{1,2} = \frac{N + c \pm \sqrt{(N - c)^2 + 4 b^2}}{2}.
\]

\[
\lambda_1 + \lambda_2 = N + c > 0,
\]

\[
\lambda_1 \lambda_2 = N c - b^2.
\]
Therefore the eigenvalues are both positive if and only if $N c - b^2>0$. From Cauchy-Schwarz inequality we know:
\[
\left( \sum_{i = t}^{N} r_i \right)^2 \leq N \sum_{i = t}^{N} r_i^2,
\]
which implies:
\(
\lambda_1 \lambda_2 = N c - b^2 \geq 0.
\)
Equality holds if and only if all \( r_i \) are equal. When at least two \( r_i \) are distinct, the inequality is strict:
\[
\lambda_1 \lambda_2 > 0.
\]

The persistent excitation condition requires:
\[
m I \leq S \leq M I,
\]
for some constants \( 0 < m \leq M \). Since \( S \) has positive eigenvalues \( \lambda_{\min} \) and \( \lambda_{\max} \), we can set:
\[
m = \lambda_{\min}, \quad M = \lambda_{\max}.
\]

The above analysis shows that as long as for any $t$, there is some $N$ such that Algorithm \ref{alg:optimizer} sets at least two distinct interest rates from time $t$ to $t+N$, then the excitation condition of paper \cite{canetti1989convergence} is met. 

In Algorithm \ref{alg:optimizer}, the rate $\rt$ is sampled from a Gaussian distribution. The mean of this distribution is the optimal rate based on the latest estimates of the parameters $\abn$, $\bbn$, $\aln$, and $\bln$, and the variance is determined by the variance of the optimal rate estimation. The covariance matrix used for this estimation is derived from $\mathbf{P}^b_t$ and $\mathbf{P}^l_t$.
It is important to note that as long as at least one of the matrices $\mathbf{P}^b_t$ or $\mathbf{P}^l_t$ has a positive element in its main diagonal (diameter), the rate $\rt$ is sampled from a distribution with non-zero variance. This implies that with high probability, after some of trials, the sampled $\rt$ will differ, even in the worst-case scenario. To determine the required number of timeslots $N$ to get at least two different $\rt$ with probability at least $1-\delta$ we conduct the following analysis:

\textbf{Claim.} \\
Let $X$ be a random variable following a Gaussian distribution with mean $\mu$ and variance $\zeta$ (i.e., $X \sim \mathcal{N}(\mu, \zeta)$). Let $\delta \in (0,1)$ be a desired confidence level. Then, the minimum number of independent samples $N$ needed to ensure that, with probability at least $1 - \delta$, there are at least two different samples among the $N$ samples is given by:

\[
N \geq \frac{\ln(\delta)}{\ln\left( \dfrac{\epsilon}{\sqrt{2\pi \zeta}} \right)},
\]

where $\epsilon$ is the quantization level (measurement precision).

\textbf{Proof of the claim}

The maximum value of the PDF occurs at the mean $\mu$:
\(
f_{\text{max}} = f(\mu) = \dfrac{1}{\sqrt{2\pi \zeta}}
\). The maximum probability that a sample falls into any single bin (the bin containing $\mu$) is approximately:
\(
p_{\text{max}} = f_{\text{max}} \times \epsilon = \dfrac{\epsilon}{\sqrt{2\pi \zeta}}.
\) Assuming independence, the probability that all $N$ samples fall into the same bin is:
\[
P(\text{all samples in the same bin}) = p_{\text{max}}^{\, N}.
\]

We require:
\[
1 - P(\text{all samples in the same bin}) \geq 1 - \delta \implies P(\text{all samples in the same bin}) \leq \delta
\]

\[
p_{\text{max}}^{\, N} \leq \delta \implies N \ln(p_{\text{max}}) \leq \ln(\delta)
\]

Substitute $p_{\text{max}}$:
\[
N \geq \dfrac{\ln(\delta)}{\ln\left( \dfrac{\epsilon}{\sqrt{2\pi \zeta}} \right)} =  \Theta\left(\dfrac{\ln  \delta}{\ln \epsilon - \ln\zeta }\right) = \Theta\left(\dfrac{-\ln  \delta}{-\ln \epsilon + \ln\zeta }\right)
\]

Applying the following claim to our case, $\epsilon = \Delta r$ (precision of the interest rate ), $\zeta = \Theta\left(\text{diam}(\mathbf{P}^b_t + \mathbf{P}^l_t)\right)$. Hence as long as $\text{diam}(\mathbf{P}^b_t + \mathbf{P}^l_t) > 0$, there is some bounded $N$ that satisfies the condition.

So now we can apply the results of \cite{canetti1989convergence} for the convergence of the RLS algorithm to the right $\theta_t$. They write the estimation error $\Tilde{\theta}_t=\theta_t-\hat{\theta}_t$ as sum of three terms $\Tilde{\theta}_t \coloneqq \Tilde{\theta}^1_t + \Tilde{\theta}^2_t + \Tilde{\theta}^3_t$. Given that $\theta$ has some major change at time $t_0$, the question is at time $t>t_0$ what is an upper bound of each of $\Tilde{\theta}^1_t, \Tilde{\theta}^2_t, \Tilde{\theta}^3_t$? They show that the first and second terms decay exponentially over time i.e., $\|\Tilde{\theta}^1_t + \Tilde{\theta}^2_t\| = \mathcal{O}(\ff^t)$ whereas the third term $\Tilde{\theta}_t^3$ which is the only term that is a function of the observation noise, is as follows:
\[
\limsup_{t \to \infty} \mathbb{E} \left( \|\tilde{\theta}_t^3 \|^2 \right) \leq \nu^2 \frac{M}{m^2} \left( \frac{1-\ff^N}{\rho^N} \right)^2 \frac{1}{\ln \left(\frac{1}{\rho}\right)} = \mathcal{O}\left(\left( \frac{1-\ff^N}{\rho^N} \right)^2\frac{\nu^2}{\ln{\frac{1}{\ff}}}\right)
\]
Where $\nu$ is the variance of the observation noise.

\end{proof}

%%% the original version of the excited condition in the paper results in a term (\sum x_i) x_t^T which always has an eigenvalue zero therefore it seems that term is not correct and I have seen the expression I am using here in other papers as conditions for excitedness so i guess the paper has made a mistake

\subsection{Proof of theorem \ref{theorem:rate-deviation-aave}}
\begin{proof}

\[
U = \frac{-\abn r + \bbn}{L}
\]

Case 1: \(U \leq \uoptimal\)

\[
U = \frac{-\abn  U R_{\text{slope1}}+ \bbn \uoptimal}{L\, \uoptimal}
\quad \implies 
U = \frac{\bbn}{L + \frac{\abn R_{\text{slope1}}}{\uoptimal}}
\quad \implies
r = \frac{ \bbn R_{\text{slope1}}}{\uoptimal \,L + \abn R_{\text{slope1}}}
\]

\[
r - \ropt = \frac{ \bbn R_{\text{slope1}}}{\uoptimal \,L + \abn R_{\text{slope1}}} - \frac{\bbn - L\, \uoptimal}{\abn }
\]

\[
R_{\text{slope1}} = \frac{\bbn - L\, \uoptimal}{\abn }
\]

\[
R_{\text{slope1}} \geq \frac{\bbn - \uoptimal L}{\abn}
\]

Case 2: \(U_t > \uoptimal\)

\begin{align}
U &= \frac{-\abn \left(R_{\text{slope1}} + R_{\text{slope2}} \left(\frac{U - \uoptimal}{1 - \uoptimal}\right)\right) + \bbn}{L} \implies 
U = \frac{\bbn - \abn R_{\text{slope1}} + \frac{\abn R_{\text{slope2}} \uoptimal}{1 - \uoptimal}}{L + \frac{\abn R_{\text{slope2}}}{1 - \uoptimal}} \\&\implies 
r  = R_{\text{slope1}} + \frac{R_{\text{slope2}} \left(\bbn - \abn R_{\text{slope1}} - \uoptimal L\right)}{\left(L + \frac{\abn R_{\text{slope2}}}{1 - \uoptimal}\right)(1 - \uoptimal)}
\end{align}

\[
r - \ropt = R_{\text{slope1}} + \frac{R_{\text{slope2}} \left(\bbn - \abn R_{\text{slope1}} - \uoptimal L\right)}{\left(L + \frac{\abn R_{\text{slope2}}}{1 - \uoptimal}\right)(1 - \uoptimal)} - \frac{\bbn - L\, \uoptimal}{\abn }
\]

\[
\rateDeviationN = 
\begin{cases} 
\left|R_{\text{slope1}} + \frac{R_{\text{slope2}} \left(\bbn - \abn R_{\text{slope1}} - \uoptimal L\right)}{\left(L + \frac{\abn R_{\text{slope2}}}{1 - \uoptimal}\right)(1 - \uoptimal)} - \frac{\bbn - L \, \uoptimal}{\abn }\right| & \text{if } R_{\text{slope1}} < \frac{\bbn - L \, \uoptimal}{\abn }, \\
\\
\left|\frac{ \bbn R_{\text{slope1}}}{\uoptimal \, L + \abn R_{\text{slope1}}} - \frac{\bbn - L \, \uoptimal}{\abn }\right| & \text{if } R_{\text{slope1}} > \frac{\bbn - L \, \uoptimal}{\abn }.
\end{cases}
\]

\[
\rateDeviationN = 
\begin{cases} 
\frac{\Delta}{1+\frac{\abn\cdot\rslopeOne}{L\,\uoptimal}} & \text{if } R_{\text{slope1}} < \frac{\bbn - L \, \uoptimal}{\abn }, \\
\\
\frac{\Delta}{1+\frac{\abn\cdot\rslopeTwo}{L(1-\uoptimal)}} & \text{if } R_{\text{slope1}} > \frac{\bbn - L \, \uoptimal}{\abn }.
\end{cases}
\]

\[
\rateDeviationN \geq \frac{\Delta}{1+\max\{\frac{\abn}{L}\}\cdot\max\{\frac{\rslopeOne}{\uoptimal}, \frac{\rslopeTwo}{1-\uoptimal}\}}
\]
\end{proof}

\subsection{Proof of theorem \ref{theorem:rev-max}}
\begin{proof}
Given the demand function $\bt = -\abn \rt + \bbn$ and the supply function $\lt = \aln \rt \ut - \bln$, we first find $\ut$ by solving the following equation for $\ut$:
\[\ut = \frac{-\abn \rt + \bbn}{\aln \rt \ut - \bln}\]

Solving this equation will yield: 
\begin{equation}\label{eq:app-u-as-a-function-of-r}
    \ut = \frac{\bln + \sqrt{(\bln)^2 - 4\aln \rt (\abn \rt - \bbn)}}{2\aln \rt}    
\end{equation}  %the value of $r$ that maximizes the sum of demand and supply is:
% \[
% r = \frac{b_b + \sqrt{b_b^2 + 4a_b^2 b_l^2}}{2a_b}
% \]
To maximize the sum of demand and supply, we first express the sum in terms of $r$:

\begin{align*}
\bt + \lt &= -\abn r + \bbn + \left( \frac{\aln r \bln + \aln r \sqrt{(\bln)^2 - 4 \aln r (\abn r - \bbn)}}{2 \aln r} - \bln \right) \\
&= -\abn r + \bbn + \left( \frac{\bln}{2} + \frac{\sqrt{(\bln)^2 - 4 \aln r (\abn r - \bbn)}}{2} - \bln \right) \\
&= -\abn r + \bbn - \frac{\bln}{2} + \frac{\sqrt{(\bln)^2 - 4 \aln r \,(\abn r - \bbn)}}{2}.
\end{align*}

We define:
\[
f(r) \coloneqq - \abn r + \bbn - \frac{b_l}{2} + \frac{\sqrt{(\bln)^2 - 4\aln r (\abn r - \bbn)}}{2}
\]

We differentiate $f(r)$ with respect to $r$ and set the derivative to zero to find the critical points:

\[
f'(r) = -\abn + \frac{d}{dr} \left( \frac{\sqrt{(\bln)^2 - 4\aln r (\abn r - \bbn)}}{2} \right)
\]

\[
\frac{d}{dr} \left( \sqrt{(\bln)^2- 4\aln r (\abn r - \bbn)} \right) = \frac{1}{2} \left( (\bln)^2 - 4\aln r (\abn r - \bbn) \right)^{-1/2} \cdot \frac{d}{dr} \left( (\bln)^2 - 4\aln r (\abn r - \bbn) \right)
\]

We compute the inner derivative:

\[
\frac{d}{dr} \left( (\bln)^2 - 4\aln r (\abn r - \bbn) \right) = -4\aln (\abn r - \bbn) - 4\aln r (\abn) = -4\aln (\abn r - \bbn + \abn r) = -4\aln (2\abn r - \bbn)
\]

Thus:

\[
\frac{d}{dr} \left( \sqrt{(\bln)^2 - 4\aln r (\abn r - \bbn)} \right) = \frac{-2\aln (2\abn r - \bbn)}{\sqrt{(\bln)^2 - 4\aln r (\abn r - \bbn)}}
\]

Therefore:

\[
f'(r) = -\abn - \frac{\aln (2\abn r - \bbn)}{2 \sqrt{(\bln)^2 - 4\aln r (\abn r - \bbn)}}
\]

Set $f'(r) = 0$ and solve for $r$:

\[
2\abn \sqrt{(\bln)^2 - 4\aln r (\abn r - \bbn)} = -\aln (2\abn r - \bbn)
\]

Square both sides and expand and simplify:

\[
4(\abn)^2 (\bln)^2 - 16(\abn)^2 \aln r (\abn r - \bbn) = (\aln)^2 (4(\abn)^2 r^2 - 4\abn r \bbn + (\bbn)^2) \implies
\]

\begin{equation}\label{eq:app-proof-opt-r}
   r = \frac{\bbn + \sqrt{(\bbn)^2 + 4(\abn)^2 (\bln)^2}}{2\abn} 
\end{equation}

Due to the constraints of the optimization problem $\ut < \umax$, if the above $\rt$ results in a utilization higher than $\umax$ we cannot set that. In order to make sure this constraint is met we plug $r$ from Equation \ref{eq:app-proof-opt-r} in Equation \ref{eq:app-u-as-a-function-of-r} and find the resulting $\ut$ which is $\frac{\abn\,\bln (1 + \sqrt{1 - 4\, \abn \, \aln})}{\aln \left( \bbn + \sqrt{(\bbn)^2 + 4\, (\abn\,\bln)^2} \right)}$. Hence if $\frac{\abn\,\bln (1 + \sqrt{1 - 4\, \abn \, \aln})}{\aln \left( \bbn + \sqrt{(\bbn)^2 + 4\, (\abn\,\bln)^2} \right)} > \umax$, the interest rate should be set such that $\ut = \umax$ therefore:

\begin{equation*}
    \umax = \frac{-\abn \rt + \bbn}{\aln \rt \umax - \bln} \implies \rt =  \frac{\bbn + \bln \umax}{\abn + \aln (\umax)^2} 
\end{equation*}

\end{proof}

\subsection{Proof of Proposition \ref{theorem:dynamic-incentive}}
\begin{proof}
    Consider a user $i$ with negligible $\bt(i)$, no matter what $i$ reports $\hat{f}(r;\theta)$ formed by the protocol remains the same hence $i$ cannot change the interest rate in any way. $i$ chooses $\hat{B}_t(i)$ to maximizes his own utility function for a given $\rt$:
    \[ -\hat{B}_t(i)\,r_t - \left(B_t(i) - \hat{B}_t(i)\right) \,r^b_t(i)\]
    The maximum points of this function are $\hat{B}_t(i) = \bt$ if $\rt < r^b_t(i)$ and $\hat{B}_t(i)=0$ otherwise. This is by definition the truthful strategy. A similar argument holds for a lender with negligible $\lt(i)$.
    
\end{proof}

\subsection{Proof of Theorem \ref{theorem:rate-manipulation}}
\begin{proof}
First, we only focus on the case that there is one strategic borrower without any strategic lender and then we show how this result changes in the presence of a strategic lender. 

\noindent\textbf{Strategic borrower} We outline an adversarial strategy that a major borrower denoted by $\advb$ might employ to mislead the protocol into selecting a lower interest rate. Consider \advb controlling $\delta_b$ fraction of the total demand $\bbn$, with a private interest rate $r^b$. In the presence of such an adversary, the true demand curve, alongside the continuum users' linear function, is depicted as the black curve $B(r)$ in Figure \ref{fig:proof-intuition}. The true preference of \advb is to borrow $\bbn \delta_b$ whenever $r < r^b$. 

Ideally during the exploration phase \ai will learn the demand curve $B(r)$ and during the exploitation phase, selects the optimal interest rate $\ropt$ such that $\ropt = \arg\min_r \left| \frac{B(r)}{L} - \uoptimal \right|$.

However, the adversary can misreport their value by pretending that the maximum rate they are willing to pay is $\bar{r}^b$, by repaying whenever the interest rate exceeds $\bar{r}^b$ during the exploration phase. Consequently, \advb deceives \ai into believing that the red curve $\bar{B}(r)$ in Figure \ref{fig:proof-intuition} represents the true demand curve. This misreporting leads \ai to choose a different interest rate, $\bar{r} = \arg\min_r \left| \frac{\bar{B}(r)}{L} - \uoptimal \right|$.

This proof addresses two key questions: 
1) When is it beneficial for the adversary to misreport their value considering that during the exploration phase they might financially suffer from lying?
2) Given that misreporting is advantageous, to what extent can the adversary influence the interest rate?

\begin{figure}[H]
    \centering
\includegraphics[width=0.9\textwidth]{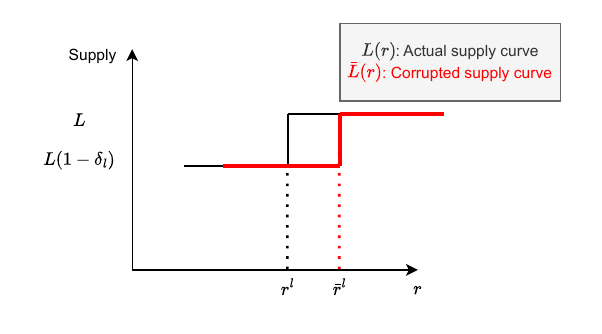}
    \caption{}
    \label{fig:proof-intuition}
\end{figure}

We start with the first point, We first compare the utility of an honest strategy and that of the adversarial strategy doe the major borrower during the exploration and exploitation phases of \ai.

\begin{align}
    \text{Utility}_b^{\text{honest}} \geq & -\Bigg(\delta_b \bbn \tau \zeta \frac{r^b - \rminb}{\rmaxl - \rminb} \frac{r^b + \rminb}{2} + C_{\tau, \zeta}^{r^b} + \delta_b \bbn (T - \tau) r^b \Bigg)
\end{align}

If the borrower follows the honest strategy and reports $r^b$, during the exploration phase, they will borrow whenever $r \in [\rminb, r^b]$ and pay an average interest rate of $\frac{\rminb + r^b}{2}$ during those timeslots. There are $\tau \zeta \frac{r^b - \rminb}{\rmaxl - \rminb}$ such timeslots based on the uniform randomization of \ai described in \prettyref{app:lc_auction}. During the rest of the exploration phase, i.e., when $r > r^b$ or during the non-randomized times, we denote the cost of the borrower by $C_{\tau, \zeta}^{r^b}$. Finally, during the exploitation phase $T - \tau$, the borrower pays at most rate $r^b$ (potentially even lower depending on the rest of the parameters and objective utilization $\uoptimal$).

Now, we discuss the utility of the borrower by misreporting $\bar{r}^b < r^b$ instead of $r^b$:

\begin{align}
    \text{Utility}_b^{\text{adv}} \leq & -\Bigg(\delta_b \bbn \tau \zeta \frac{\bar{r}^b - \rminb}{\rmaxl - \rminb} \frac{\bar{r}^b + \rminb}{2} + \delta_b \bbn \tau \zeta \frac{r^b - \bar{r}^b}{\rmaxl - \rminb} r^b \nonumber \\
    & + C_{\tau, \zeta}^{r^b} + \delta_b \bbn (T - \tau) \bar{r}^b \Bigg)
\end{align}

During the randomization phase, whenever $r < \bar{r}^b$, the adversarial borrower pays $r$. There are $\tau \zeta \frac{\bar{r}^b - \rminb}{\rmaxl - \rminb}$ such timeslots with an average rate $\frac{\bar{r}^b + \rminb}{2}$. However, the adversary incurs a loss compared to the honest strategy during the exploration phase when $\bar{r}^b < r < r^b$, because they cannot borrow from the protocol in these timeslots, even though following the honest strategy, they could have borrowed and made a strict profit compared to borrowing from outside at rate $r^b$. This extra cost is the second term $\delta_b \bbn \tau \zeta \frac{r^b - \bar{r}^b}{\rmaxl - \rminb} r^b$. During the rest of the exploration phase timeslots, the cost is the same as if they had followed the honest strategy, hence $C_{\tau, \zeta}^{r^b}$. Finally, during the exploitation phase, the adversary at best forces \ai to set $\bar{r}^b$ (we consider $\bar{r}^b$ to be the minimum enforceable rate by \advb). Hence by solving the trade-off between the extra cost incurred during the exploration phase and the strict profit made during the exploration phase bu following the adversarial strategy, we find an upper bound on the set of profitable $\bar{r}^b$.

\begin{align}
    \text{Utility}_b^{\text{adv}} - \text{Utility}_b^{\text{honest}} \leq & -\delta_b \bbn \Bigg( \frac{\tau \zeta}{2} \frac{(r^b - \bar{r}^b)^2}{\rmaxl - \rminb} + \big(T - \tau\big) (\bar{r}^b - r^b) \Bigg)
\end{align}

\begin{align}
    & \text{Utility}_b^{\text{adv}} - \text{Utility}_b^{\text{honest}} \geq 0 \\
    & \implies \Bigg( \frac{\tau \zeta}{2} \frac{(r^b - \bar{r}^b)^2}{\rmaxl - \rminb} + \big(T - \tau\big) (\bar{r}^b - r^b) \Bigg) \leq 0 \\
    & \implies 0 < r^b - \bar{r}^b < \frac{2(\rmaxl - \rminb)}{\zeta} \left( \frac{T}{\tau} - 1 \right)\label{eq:app-tau-bound-for-r}
\end{align}

Next, we address the second point: given that \ai aims to minimize $|\bar{B}(r) - \uoptimal|$, what is the minimum $\bar{r}^b$ that \advb can choose to force \ai to set $\bar{r} = \bar{r}^b$? Additionally, how far can $\bar{r}$ deviate from $\ropt$?

We first express $\ropt$ as a function of the remaining parameters. If the adversary reports $r^b$, \ai solves:

\[
\frac{\bbn(1 - \delta_b) - \abn \, r}{L} = \uoptimal \implies r = \frac{\bbn(1 - \delta_b) - L \, \uoptimal}{\abn}
\]

If $r > r^b$, \ai does not require the adversary's demand to achieve optimal utilization. Hence, regardless of the interest rate reported by the adversary, \ai sets the steady-state interest rate to $\frac{\bbn(1 - \delta_b) - L \, \uoptimal}{\abn}$, resulting in no manipulation ($\ropt - \bar{r} = 0$).

However, if $\frac{\bbn(1 - \delta_b) - \abn \, r^b}{L} < \uoptimal$, the utilization with the demand generated by infinitesimal borrowers is lower than $\uoptimal$ at $r^b$. Therefore, \ai might need to set a lower interest rate than $r^b$ to incorporate the adversary's demand. Several scenarios arise:

\begin{itemize}
    \item If $\frac{\bbn - \abn \, r^b}{L} < \uoptimal$, even including the adversary's demand at $r^b$, the utilization remains low. In this case, $\ropt$ will be lower than $r^b$ to attract further demand, given by $\ropt = \frac{\bbn - L \, \uoptimal}{\abn} < r^b$. The adversary can report any $\bar{r}^b$ that satisfies the following conditions to convince \ai to select $\bar{r}^b$:

    \begin{equation}\label{eq:app-borrower-rate}
        \frac{\bbn(1 - \delta_b) - \abn \, \bar{r}}{L} < \uoptimal \implies \bar{r} > \frac{\bbn(1 - \delta_b) - L \, \uoptimal}{\abn}
    \end{equation}

    and

    \begin{equation}\label{eq:app-borrower-rate-1}
        \uoptimal - \frac{\bbn(1 - \delta_b) - \abn \, \bar{r}}{L} > \frac{\bbn - \abn \, \bar{r}}{L} - \uoptimal \implies \bar{r} > \frac{\bbn(1 - \frac{\delta_b}{2}) - L \, \uoptimal}{\abn}
    \end{equation}

    In this scenario:

    \[
    |\ropt - \bar{r}| = \left|\frac{\bbn - L \, \uoptimal}{\abn} - \frac{\bbn(1 - \frac{\delta_b}{2}) - L \, \uoptimal}{\abn}\right| = \frac{\bbn \delta_b}{2 \abn}
    \]

    \item Next, we consider the following scenario: $\frac{\bbn - \abn \, r^b}{L} > \uoptimal$ and $\frac{\bbn - \abn \, r^b}{L} - \uoptimal < \uoptimal - \frac{\bbn(1 - \delta_b) - \abn \, r^b}{L}$. In this case, attracting the adversary's demand will cause the utilization to exceed $\uoptimal$, but it is closer to $\uoptimal$ than the case where the adversary is not involved. Consequently, \ai sets $\ropt = r^b - \varepsilon$. The adversary can only choose $\bar{r}$ that satisfies conditions \eqref{eq:app-borrower-rate} and \eqref{eq:app-borrower-rate-1}; otherwise, \ai will choose $r > \bar{r}$. Therefore, we have:

\[
\bar{r} > \frac{\bbn(1 - \frac{\delta_b}{2}) - L \, \uoptimal}{\abn}
\]

Moreover, we know in this case $\ropt = r^b \leq \frac{\bbn - L \, \uoptimal}{\abn}$. Hence, we again find:

\[
|\ropt - \bar{r}| \leq \frac{\bbn \delta_b}{2 \abn}
\]

    \item Now, consider the scenario: $\frac{\bbn - \abn \, r^b}{L} > \uoptimal$ and $\frac{\bbn - \abn \, r^b}{L} - \uoptimal > \uoptimal - \frac{\bbn(1 - \delta_b) - \abn \, r^b}{L}$. In this case, attracting the adversary's demand will cause the utilization to exceed $\uoptimal$, and the protocol achieves a utilization closer to $\uoptimal$ without the adversary's demand. Consequently, the protocol sets $\ropt = r^b + \varepsilon$.
In this scenario, the adversarial borrower cannot change the steady-state interest rate by misreporting a lower interest rate because the protocol is not willing to set any interest rate equal to or lower than $r^b$ since this increases the utilization excessively. Therefore, $\ropt - \bar{r} = 0$.
\end{itemize}

\noindent\textbf{Strategic lender} Now, we consider the scenario where there is only one strategic lender, denoted as \advl, without any strategic borrowers. The adversarial strategy for the lender is to report a higher rate $\bar{r}^l$ instead of $r^l$. Refer to Figure \ref{fig:proof-intuition-lender} for an illustration.

\begin{figure}[H]
    \centering
\includegraphics[width=0.9\textwidth]{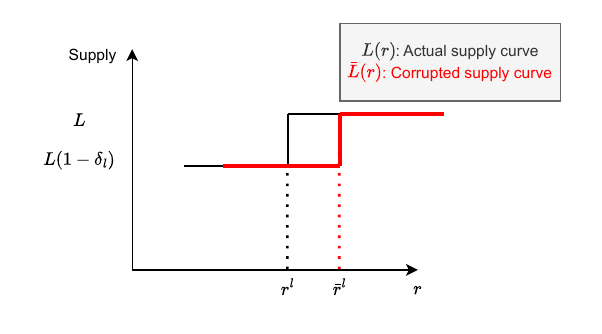}
    \caption{}
    \label{fig:proof-intuition-lender}
\end{figure} 

First, we compare the utility of \advl under the honest strategy and the adversarial strategy. We assume the lender's utility is directly proportional to $r$ for simplification:

\begin{align}
    \text{Utility}_l^{\text{honest}} \geq & \Bigg(\delta_l L \tau \zeta \frac{\rmaxl - r^l}{\rmaxl - \rminb} \frac{r^l + \rmaxl}{2} + C_{\tau, \zeta}^{r^l} + \delta_l L (T - \tau) r^l \Bigg)
\end{align}

\begin{align}
    \text{Utility}_l^{\text{adv}} \leq & \Bigg(\delta_l L \tau \zeta \frac{\rmaxl - \bar{r}^l}{\rmaxl - \rminb} \frac{\bar{r}^l + \rmaxl}{2} + \delta_l L \tau \zeta \frac{r^l - \bar{r}^l}{\rmaxl - \rminb} r^l \nonumber \\
    & + C_{\tau, \zeta}^{r^l} + \delta_l L (T - \tau) \bar{r}^l \Bigg)
\end{align}

By finding the range of $\bar{r}^l$ that makes $\text{Utility}_l^{\text{adv}} - \text{Utility}_l^{\text{honest}} > 0$, we can derive the exact bound as in Equation \ref{eq:app-tau-bound-for-r}.

Next, we determine $\ropt$. The controller \ai solves $\frac{\bbn - \abn\,r}{L(1 - \delta_l)} = \uoptimal$. If $r = \frac{\bbn - L(1 - \delta_l) \uoptimal}{\abn} < r^l$, the protocol can achieve the desired utilization without \advl's supply, and \advl cannot manipulate the interest rate, resulting in $\bar{r} - \ropt = 0$. If $\frac{\bbn - \abn\,r^l}{L(1 - \delta_l)} > \uoptimal$, \advl's supply is necessary for maintaining the desired utilization. The following cases are possible:

\begin{itemize}
    \item If $\frac{\bbn - \abn\,r^l}{L} > \uoptimal$, even with \advl's funds, $\ropt = \frac{\bbn - L \uoptimal}{\abn}$. In this case, \advl can misreport any $\bar{r}^l$ that satisfies:

    \begin{equation}\label{eq:app-lender-rate}
        \frac{\bbn - \abn \, \bar{r}}{L(1 - \delta_l)} > \uoptimal \implies \bar{r} < \frac{\bbn - L(1 - \delta_l) \, \uoptimal}{\abn}
    \end{equation}

    and

    \begin{equation}\label{eq:app-lender-rate-1}
        \frac{\bbn - \abn \, \bar{r}}{L(1 - \delta_l)} - \uoptimal > \uoptimal - \frac{\bbn - \abn \, \bar{r}}{L} \implies \bar{r} < \frac{\bbn - \uoptimal \, L \frac{(1 - \delta_l)}{(1 - \frac{\delta_l}{2})}}{\abn}
    \end{equation}

    Therefore, $|\bar{r} - \ropt| < \frac{\uoptimal \, L \delta_l}{\abn (2 - \delta_l)}$.

    \item If $\frac{\bbn - \abn\,r^l}{L} < \uoptimal$, but 
    \(
        \frac{\bbn - \abn \, r^l}{L(1 - \delta_l)} - \uoptimal > \uoptimal - \frac{\bbn - \abn \, r^l}{L}
    \) holds, \ai sets $r^l + \varepsilon$ to get \advl's demand with maximum utilization possible. Again, \advl can report interest rates as high as $\frac{\bbn - \uoptimal \, L \frac{(1 - \delta_l)}{(1 - \frac{\delta_l}{2})}}{\abn}$ and get \ai to set this rate. Since $\ropt = r^l > \frac{\bbn - L \uoptimal}{\abn}$, $|\bar{r} - \ropt| < \frac{\uoptimal \, L \delta_l}{\abn (2 - \delta_l)}$.

    \item If $\frac{\bbn - \abn\,r^l}{L} < \uoptimal$ and
    \(
        \frac{\bbn - \abn \, r^l}{L(1 - \delta_l)} - \uoptimal > \uoptimal - \frac{\bbn - \abn \, r^l}{L}
    \) does not hold, \ai gets closer to $\uoptimal$ by setting $\ropt = r^l - \varepsilon$ and not involving \advl's funds at all. In this case, \advl cannot manipulate the interest rate.
\end{itemize}

\noindent\textbf{Strategic Lender and Borrower} 
In the presence of both strategic lenders and borrowers, the potential manipulation of the interest rate is bounded by the influence each could exert in isolation. The strategic borrower (\advb) aims to report the minimum rate, while the strategic lender (\advl) aims to report the maximum rate. If the protocol requires only one of them to achieve the desired utilization, the scenario reduces to the case of a single adversary.
However, if \ai needs both \advb's demand and \advl's supply to achieve the desired utilization, the party controlling a larger market share may influence the rate more significantly. Despite this, the rate manipulation cannot exceed the influence they would have if they were the sole adversary, assuming all other participants are truthful. This is because \ai aims to satisfy both \advb and \advl, making it more challenging for either party to manipulate the rate to an extreme due to the inherent competition between them.

\end{proof}

\subsection{Proof of Theorem \ref{theorem:static-curve-manipulation}}
\begin{proof}
\(\advl\) has a more severe impact on the interest rate when \(\advb\) is absent, and vice versa, because they push the rate in opposite directions. Therefore, evaluating each adversary separately while assuming the other is truthful provides an upper bound on their combined adversarial effects.

We begin by analyzing the lender's utility function to determine the optimal \(\hat{L}\) that maximizes utility. For simplicity, we omit the utilization term multiplier \(\frac{B}{(1-\delta_l)L+\hat{L}}\). Moreover, we replace \(\phi(U)\) in the Utility function with the steeper section of the baseline curve, for simplicity, we call the slope and intercept of the curve in the steeper part respectively by $\alpha$ and $\beta$:
    \[
    \alpha \coloneqq \frac{\rslopeTwo}{1-\uoptimal}, \quad \beta \coloneqq \rslopeOne - \rslopeTwo\frac{\uoptimal}{1-\uoptimal}.
    \]
Focusing on the steeper part of the curve intensifies the attack, as it increases lenders' incentives to strategically withhold deposits. Even small withholdings can significantly impact the interest rate since in this region the interest rate is very sensitive to changes of utilization.

We take the derivative of utilization with respect to \(\hat{L}\) and set it to zero, to find the optimal \(\hat{L}\) for a strategic lender.

\begin{align*}
    \frac{d\, \text{Utility}_{\advl}(\hat{L})}{d\,\hat{L}} = 0 &\implies \hat{L}_{\text{strtgc}} = \frac{L(1-\delta_l)(\beta-\rolt) + \sqrt{BL(1-\delta_l)\alpha(\rolt-\beta)}}{{\rolt-\beta}} \\&= -L(1-\delta_l) + \sqrt{\frac{BL(1-\delta_l)\alpha}{\rolt-\beta}}
\end{align*}
We note that in order for the attack to be effective $\hat{L}_{\text{strtgc}}$ should be greater than zero and less than $\delta L$, hence:
\begin{equation}\label{eq:rolt-upperbound}
    \hat{L}_{\text{strtgc}} > 0\ \implies \rolt \leq \frac{B\alpha}{L(1-\delta_l)}+\beta
\end{equation}
Now we compare the utilization when $\advl$ behaves truthfully versus when it behaves strategically. Behaving truthfully based on the definition means  that $\advl$ will add deposit as long as the $\phi(\frac{B}{(1-\delta)L+\hat{L}})\geq \rolt$, so the truthful supply for \advl is : \[
\hat{L}_{\text{trthfl}} \coloneqq \arg\max_{\hat{L} \leq \delta L} \left\{ \hat{L} \, \middle| \, \alpha\left(\frac{B}{(1-\delta)L + \hat{L}}\right) +\beta\geq \rolt \right\}
\]
We are interested in $\max |U_{\text{strtgc}}-U_{\text{trthfl}}|$ where $U_{\text{strtgc}} \coloneqq \frac{B}{L(1-\delta_l)+\hat{L}_{\text{strtgc}}}$ and $U_{\text{trthfl}} \coloneqq \frac{B}{L(1-\delta_l)+\hat{L}_{\text{trthfl}}}$

We have:
\begin{align*}
    |U_{\text{strtgc}}-U_{\text{trthfl}}| &= U_{\text{strtgc}}-U_{\text{trthfl}} \\&\underset{\hat{L}_{\text{trthfl}} \leq \delta_l L}{\leq} U_{\text{strtgc}}-\frac{B}{L} \\ 
    &\underset{\ref{eq:rolt-upperbound}}{\leq} \frac{B}{L(1-\delta_l)} - \frac{B}{L} \\&\leq \frac{\delta_l B}{L(1-\delta_l)}
\end{align*}

Hence:

\begin{align*}
    \advImpactNs &= |r_{\text{strtgc}} - r_{\text{trthfl}}| \\&= \alpha   |U_{\text{strtgc}}-U_{\text{trthfl}}|\\ 
    &\leq \frac{B\delta_l \rslopeTwo}{L(1-\delta_l)(1-\uoptimal)}
\end{align*}

Now we do a similar analysis for the borrower:

\begin{align}\label{eq:bstartegic}
    \frac{d\, \text{Utility}_{\advb}(\hat{B})}{d\,\hat{B}} = 0 &\implies \hat{B}_{\text{strtgc}} = \frac{\robt-\beta}{\frac{2\alpha}{L}} - \frac{B(1-\delta_b)}{2}
\end{align}

Moreover, by definition:
\[
\hat{B}_{\text{trthfl}} \coloneqq \arg\max_{\hat{B} \leq B\delta_b } \left\{ \hat{B} \, \middle| \, \alpha\frac{B(1-\delta_b)+\hat{B}}{L} +\beta\leq \robt \right\}
\]
We note that strategic withholding attack is only relevant if the strategic player borrows strictly less than the truthful borrower therefore $\hat{B}_{\text{trthfl}}>0$, this means that \begin{equation}\label{eq:robt-lowerbound}
    \frac{\alpha B(1-\delta_b)}{L} + \beta \leq \robt
\end{equation}

We are interested in $\max |U_{\text{strtgc}}-U_{\text{trthfl}}|$ where $U_{\text{strtgc}} \coloneqq \frac{B(1-\delta_b)+\hat{B}_{\text{strtgc}}}{L}$ and $U_{\text{trthfl}} \coloneqq \frac{B(1-\delta_b)+\hat{B}_{\text{trthfl}}}{L}$. In an effective attack, \advb borrowers less compare to the truthful alternative, hence causing a lower utilization:
\begin{align*}
    |U_{\text{strtgc}}-U_{\text{trthfl}}| &= U_{\text{trthfl}} - U_{\text{strtgc}}\\&\leq \frac{B}{L} - U_{\text{strtgc}} \\&\underset{\ref{eq:bstartegic}}{=}
    \frac{B}{L} - \frac{\frac{B(1-\delta_b)}{2}+ \frac{\robt-\beta}{\frac{2\alpha}{L}}}{L} \\&\underset{\ref{eq:robt-lowerbound}}{\leq}
    \frac{B(1+\delta_b)}{2L} - \frac{ \frac{\alpha B(1-\delta_b)}{L(2\alpha / L)} }{L} = \frac{B\delta_b}{L}
\end{align*}

Hence:

\begin{align*}
    \advImpactNs &= |r_{\text{strtgc}} - r_{\text{trthfl}}| \\&= \alpha   |U_{\text{strtgc}}-U_{\text{trthfl}}|\\ 
    &\leq \frac{ B\delta_b \rslopeTwo}{L (1-\uoptimal)}
\end{align*}

\section{Least square estimation related work}

\noindent\textbf{Recursive least square estimation}
Recursive least squares (RLS) estimation is an online adaptation of the standard least squares method, enabling efficient updating of parameter estimates as new data becomes available. While applying the full least squares estimator to $n$ datapoints incurs a computational cost of $\Theta(n^2)$, the RLS algorithm reduces this complexity to $\Theta(n)$ \cite{simon2006optimal}. To accommodate time-varying parameters, the RLS algorithm is typically augmented with a forgetting factor, which assigns exponentially decaying weights to older datapoints \cite{islam2019recursive}. Notably, there exists a close relationship between RLS with a forgetting factor and Kalman filtering. In fact, \cite{sayed1994state} demonstrates that minimizing a weighted least squares loss is equivalent to estimating the state in a Kalman filter with state evolution given by
\(
x_{i+1} = \frac{x_i}{\sqrt{\lambda}},
\)
hence a smaller forgetting factor $\lambda$ is appropriate for more rapidly changing conditions.
Moreover, Canetti et al. \cite{canetti1989convergence} establish a trade-off between estimation precision and adaptivity: while setting $\lambda=1$ (i.e., no forgetting) perfectly cancels the effect of measurement noise, it impedes the algorithm's ability to track changes in the system. Therefore, selecting the appropriate forgetting factor is critical and should reflect the pace of the underlying dynamics. To this end, Paleologu et al. \cite{paleologu2008robust} propose tuning the forgetting factor based on the posterior estimation error, aligning it with the noise variance to ensure that the estimator's error matches the expected noise level.\\
\noindent\textbf{Robust linear estimation} Conventional recursive least squares (RLS) estimators lack robustness to outliers due to their linear structure. While nonlinear transformations of residuals can mitigate impulsive noise \cite{zou2000recursive,kovavcevic2016robust}, such approaches remain vulnerable to strategic adversarial corruptions that subtly bias estimates without extreme outliers. Bhati et al. \cite{bhatia2015robust} employ hard thresholding to identify adversarial samples, guaranteeing exact parameter recovery under bounded corruption. However, their framework requires multiple full-batch iterations to find the clean data points, rendering them unsuitable and inefficient for recursive settings with dynamically evolving parameters.
Chen et al. explore the adversarial robustness of Kalman filtering under probabilistic corruption models \cite{chen2022kalman}, where they theoretically guarantee sublinear regret in competing with an oracle aware of corruption locations, but they fail to suggest a practical algorithm. Alternative approaches \cite{ting2007learning} model the adversary as long-tail observation noise but with a known distribution which limits real-world applicability.

\end{proof}

%% file: algorithms/estimator.tex
\begin{algorithm}
\begin{algorithmic}\caption{Estimating the demand and supply curve using RLS with forgetting factor}
\label{alg:rls-forgetting-factor}
    \State Initialize: \(\hat{\boldsymbol{\theta}}^l_0, \hat{\boldsymbol{\theta}}^b_0\), \(\mathbf{P}^l_0, \mathbf{P}^b_0 \gets\) large positive definite matrix, \(\ff\) (forgetting factor), \(\nu\) noise standard deviation
    
    \For{each time step \( t \)}
    {\State Observe \(\rS{t-1}, \uS{t-1}\) and \( \lt, \bt \); And set $\mathbf{x}_t^b = [\rS{t-1} , 1]^T$ , $\mathbf{x}_t^l = [\rS{t-1}\uS{t-1}, 1]^T$
        \State Compute the gain vector for the supply and demand observation : 
        \[
            \mathbf{K}_t^l = \frac{\mathbf{P}^l_{t-1} \mathbf{x}^l_t}{(\ff + (\mathbf{x}^l_t)^T \mathbf{P}^l_{t-1} \mathbf{x}^l_t)}, \quad \mathbf{K}_t^b = \frac{\mathbf{P}^b_{t-1} \mathbf{x}^b_t}{(\ff + (\mathbf{x}^b_t)^T \mathbf{P}^b_{t-1} \mathbf{x}^b_t)}
        \]
        \State Update the parameter estimate: 
        \[
            \hat{\boldsymbol{\theta}}_t^l = \hat{\boldsymbol{\theta}}_{t-1}^l + \mathbf{K}^l_t (\lt - (\mathbf{x}_t^l)^T \hat{\boldsymbol{\theta}}_{t-1}^l), \quad \hat{\boldsymbol{\theta}}_t^b = \hat{\boldsymbol{\theta}}_{t-1}^b + \mathbf{K}^b_t (\bt - (\mathbf{x}_t^b)^T \hat{\boldsymbol{\theta}}_{t-1}^b)
        \]
        \State Update the covariance matrix: 
        \[
            \mathbf{P}_t^l = \frac{1}{\ff}( \mathbf{I} - \mathbf{K}_t^l (\mathbf{x}_t^l)^T ) \mathbf{P}^l_{t-1}, \quad \mathbf{P}_t^b = \frac{1}{\ff}( \mathbf{I} - \mathbf{K}_t^b (\mathbf{x}_t^b)^T ) \mathbf{P}^b_{t-1} 
        \]
        \State Parse \([\alhat, -\blhat] \gets \hat{\boldsymbol{\theta}}_t^l\) and \([-\abhat, \bbhat] \gets \hat{\boldsymbol{\theta}}_t^b\)
        }
\end{algorithmic}
\end{algorithm}

%% file: algorithms/optimizer.tex
\begin{algorithm}[t]
\caption{Setting the optimal interest rate to achieve desired utilization}
\label{alg:optimizer}
\begin{algorithmic}
    \State Initialize: \(\uoptimal \), \(\zeta\) (randomness probability), \(\xi\) (error threshold to do randomization)
    
\For{each time step $ t $}{
        \State Read $\abhat, \bbhat, \alhat, \blhat, \mathbf{P}^l_{t}, \mathbf{P}^b_{t}$ from Algorithm \ref{alg:rls-forgetting-factor}
        \State $\text{Var}(\abhat) \gets \mathbf{P}^b_{t}(1,1)$, $\text{Var}(\bbhat) \gets \mathbf{P}^b_{t}(2,2)$, $\text{Var}(\alhat) \gets \mathbf{P}^l_{t}(1,1)$, $\text{Var}(\blhat) \gets \mathbf{P}^l_{t}(2,2)$
        \State $\E{\rt} \gets \frac{\bbhat + \blhat \uoptimal}{\abhat + \alhat (\uoptimal)^2}$\label{alg-line:optimal-rate}
        \State $\text{Var}(\rt) \gets \frac{1}{\big(\abhat + \alhat (\uoptimal)^2\big)^2} \left(\text{Var}(\bbhat) + (\uoptimal)^2 \text{Var}(\blhat)\right)$
        \Statex $\quad + \frac{\big(\bbhat + \blhat \uoptimal\big)^2}{\big(\abhat + \alhat (\uoptimal)^2\big)^4} \left(\text{Var}(\abhat) + (\uoptimal)^4 \text{Var}(\alhat)\right)$
        \State Sample $\rt \sim \mathbb{N}(\E{\rt}, \text{Var}[\rt])$ \label{alg_eq:randomness}
    }
\end{algorithmic}
\end{algorithm}